%% file: draft.v2.4.tex
\pdfoutput=1
\documentclass[a4paper,fleqn,usenatbib]{mnras}

\usepackage{mathptmx}
\usepackage{txfonts}

\usepackage[T1]{fontenc}
\usepackage{ae,aecompl}

\usepackage{graphicx}	
\usepackage{amssymb}	
\usepackage{upgreek}
\usepackage{gensymb}

\newcommand{\kms}{km\,s$^{-1}$}
\renewcommand{\ion}[2]{#1$\,${\small \MakeUppercase{\romannumeral #2}}}

\title[SN 2015G]{ The Nearby Type Ibn Supernova 2015G: \\
Signatures of Asymmetry and Progenitor Constraints }

\author[Shivvers et al.]{
Isaac Shivvers,$^{1}$\thanks{E-mail: ishivvers@berkeley.edu}
WeiKang Zheng,$^{1}$
Schuyler D. Van Dyk,$^{2}$
Jon Mauerhan,$^{1}$   \newauthor
Alexei V. Filippenko,$^{1,3}$
Nathan Smith,$^{4}$
Ryan J. Foley,$^{5}$
Paolo Mazzali,$^{6,7}$   \newauthor
Atish Kamble,$^{8}$
Charles D. Kilpatrick,$^{5}$
Raffaella Margutti,$^{9}$
Heechan Yuk,$^{1,10}$   \newauthor
Melissa L. Graham,$^{1,11}$
Patrick L. Kelly,$^{1}$
Jennifer Andrews,$^{4}$
Thomas Matheson,$^{12}$   \newauthor
W. M. Wood-Vasey,$^{13}$
Kara A. Ponder,$^{13}$
Peter J. Brown,$^{14}$
Roger Chevalier,$^{15}$  \newauthor
Dan Milisavljevic,$^{8}$
Maria Drout,$^{8,16}$
Jerod Parrent,$^{8}$
Alicia Soderberg,$^{8}$   \newauthor
Chris Ashall,$^{7}$    
Andrzej Piascik,$^{7}$
Simon Prentice$^{7}$
\\
$^{1}$Department of Astronomy, University of California, Berkeley, CA 94720-3411, USA \\
$^{2}$IPAC, Caltech, Mail Code 100-22, Pasadena, CA 91125, USA \\
$^{3}$Senior Miller Fellow, Miller Institute for Basic Research in Science, University of California, Berkeley, CA 94720, USA \\
$^{4}$Steward Observatory, University of Arizona, 933 N. Cherry Ave., Tucson, AZ 85721, USA \\
$^{5}$Department of Astronomy and Astrophysics, University of California, Santa Cruz, CA 95064, USA \\
$^{6}$Max-Planck-Institut f\"{u}r Astrophysik, Karl-Schwarzschild-Strasse 1, D-85748 Garching, Germany \\
$^{7}$Astrophysics Research Institute, Liverpool John Moores University, Liverpool L3 5RF, UK \\
$^{8}$Harvard-Smithsonian Center for Astrophysics, 60 Garden St., Cambridge, MA 02138, USA \\
$^{9}$CIERA, Department of Physics and Astronomy, Northwestern University, Evanston, IL 60208, USA \\
$^{10}$Department of Physics and Astronomy, San Francisco State University, San Francisco, CA 94132, USA \\
$^{11}$Department of Astronomy, University of Washington, Box 351580, U.W., Seattle, WA 98195-1580, USA \\
$^{12}$National Optical Astronomy Observatory, Tucson, AZ 85719, USA \\
$^{13}$PACC, Physics and Astronomy Department, University of Pittsburgh, Pittsburgh, PA 15260, USA  \\
$^{14}$Department of Physics and Astronomy, Texas A\&M University, 4242 TAMU, College Station, TX 77843, USA \\
$^{15}$Department of Astronomy, University of Virginia, P.O. Box 400325, Charlottesville, VA 22904, USA \\
$^{16}$Carnegie Observatories, 813 Santa Barbara Street, Pasadena, CA 91101, USA \\
}

\date{Accepted for publication in MNRAS}
\pubyear{2017}
\begin{document}
\label{firstpage}
\pagerange{\pageref{firstpage}--\pageref{lastpage}}
\maketitle

\begin{abstract}

We present the results of an extensive observational campaign
on the nearby Type Ibn SN~2015G,
including data from radio through ultraviolet wavelengths.
SN~2015G was asymmetric, showing late-time nebular lines 
redshifted by $\sim 1000$\,\kms.  It shared many features with the prototypical
SN~Ibn~2006jc, including extremely strong \ion{He}{1} emission lines and
a late-time blue pseudocontinuum. The young SN~2015G showed narrow
P-Cygni profiles of \ion{He}{1}, but never in its evolution did it show any signature
of hydrogen --- arguing for a dense, ionized, and hydrogen-free circumstellar medium
moving outward with a velocity of $\sim 1000$\,\kms\ and
created by relatively recent mass loss from the progenitor star.
Ultraviolet through infrared observations show that the fading SN~2015G
(which was probably discovered some 20\,d post-peak) had a 
spectral energy distribution that was well described by a simple, single-component blackbody.
Archival {\it HST} images provide upper limits on the luminosity of SN~2015G's progenitor,
while nondetections of any luminous radio afterglow and optical nondetections of outbursts
over the past two decades provide constraints upon its mass-loss history.


\end{abstract}




\section{Introduction}

A basic understanding of core-collapse supernovae (SNe) as
luminous displays marking the collapse of a massive stellar core
has been in place for at least half a century
\citep[e.g.,][]{1966ApJ...143..626C,1971ApJ...163...11A}.
Remarkably, new observations continue to find extreme examples of the process,
some of which test the boundaries of our understanding.

For example, binarity within massive-star populations appears to produce complex and
only partially understood diversity in supernova (SN) properties via pre-explosion mass exchange 
and mass loss, likely leading to the population of stripped-envelope SNe
\citep[Types IIb/Ib/Ic; e.g.,][]{1992ApJ...391..246P,2011MNRAS.412.1522S,2012Sci...337..444S,2017PASP..129e4201S}.
Evidence is mounting from observations of interacting (Type IIn) SNe  
that their progenitors undergo extreme episodes of mass loss shortly before core collapse,
creating dense circumstellar material \citep[CSM; see review by][]{2014ARA&A..52..487S},
but exactly what mechanism is powering these death throes remains unclear.
In some cases, enhanced (eruptive) mass loss occurs only a few years to decades
before core collapse, which may point to instabilities in late
nuclear burning phases triggering mass loss or binary interaction
\citep[e.g.,][]{2012MNRAS.423L..92Q,2013MNRAS.430.1801M,2014ApJ...780...21M,2014ApJ...785...82S}.

Connecting the stripped-envelope and interacting SN populations are the 
rare Type Ibn SNe \citep[e.g.,][]{2000AJ....119.2303M,2007ApJ...657L.105F,2008MNRAS.389..113P}.
These core-collapse SNe exhibit the narrow spectral emission lines characteristic
of an ionized CSM and other key indications of dense CSM \citep[e.g.,][]{2009MNRAS.400..866C};
however, spectra of SNe~Ibn show little or no hydrogen emission and instead are dominated by
strong helium emission lines (most notably \ion{He}{1} $\lambda\lambda$5876, 6678, and 7065).
Weak-hydrogen examples also exist as intermediate Type IIn/Ibn SNe
\citep[e.g.,][]{2008MNRAS.389..131P,2012MNRAS.426.1905S,2015MNRAS.449.1921P},
and there are a few known examples of hydrogen-weak explosions (Type Ib SNe)
which then interacted with shells of hydrogen-rich
material lost by the progenitor tens to thousands of years before core collapse
\citep[e.g., SNe~2001em and 2014C;][]{2006ApJ...641.1051C,2015ApJ...815..120M,2017ApJ...835..140M}.

Only about 25 SNe~Ibn are known at this time, and the properties of
this subclass are just beginning to be mapped out
\citep[e.g.,][]{2013ApJ...769...39S,2016MNRAS.456..853P,2016MNRAS.461.3057S,2017ApJ...836..158H}.
SN~2015G, which exploded in the outskirts of
NGC~6951 at a distance of 23.2\,Mpc, is one of the nearest known SN~Ibn to date;
as such, it has allowed us the opportunity to study a member
of this rare subclass in detail. 
NGC~6951 has also hosted two other SNe in the past few decades: SN~IIn~1999el \citep{2002ApJ...573..144D}
and SN~Ia~2000E \citep{2001A&A...372..824V,2003ApJ...595..779V}.
In this paper, we present the results of an extensive observational campaign on SN~2015G,
from radio wavelengths to the ultraviolet (UV) and spanning nearly a year of follow-up observations.
In \S\ref{sec:observations} we present our observations, in \S\ref{sec:discussion} we put those
data into context and calculate the implied physical properties of the system, and
in \S\ref{sec:conclusion} we summarise and conclude.

\section{Observations}
\label{sec:observations}

SN~2015G was discovered by Kunihiro Shima at 15.5\,mag (unfiltered)
on 2015-03-23.778 (we use UT dates and times throughout this article)
and spectroscopically classified as a SN~Ibn, similar to SN~2006jc, 3\,d 
afterward \citep{2015CBET.4087....1Y,2015ATel.7298....1F}.
We initiated a photometric and spectroscopic follow-up
effort for SN~2015G as soon as
its nature as a nearby example of the rare Type Ibn subclass was understood.
From the ground, this campaign included a regular cadence of imaging through {\it BVRI} filters, a detailed
spectroscopic follow-up campaign at both low and moderate resolution, two epochs of 
near-IR imaging ({\it J, H,} and $K_s$ filters), and three epochs of radio-wavelength observations.
We also obtained multiple epochs of space-based UV imaging with {\it Swift} and
with the {\it Hubble Space Telescope (HST)},
three epochs of {\it HST} UV spectroscopy, and two epochs of {\it HST} optical imaging.
Though some of these observations produced only
nondetections, the combination of detections and upper limits 
forms an extensive dataset on SN~2015G.

Unfortunately, we did not catch SN~2015G before it reached peak brightness ---
comparisons between the early unfiltered amateur photometry and ours
(beginning 4\,d later) indicate that the SN was already on the decline at the time of discovery.
Because the peak was unobserved, throughout this paper we refer to the time since
the {\it discovery date} as the phase of SN~2015G.

\subsection{Ultraviolet through Infrared Imaging}
\label{sec:imaging}

Filtered {\it BVRI} and unfiltered observations of SN~2015G were obtained with
the 0.76\,m Katzman Automatic Imaging Telescope \citep[KAIT;][]{2001ASPC..246..121F}
at Lick Observatory nearly nightly from days 4 through 37.  
As the SN faded below KAIT's detection threshold, we began a campaign
with the Direct Imaging Camera on the Lick 1\,m Nickel telescope.
We maintained a regular observing cadence until day 155,
at which point SN~2015G faded below our
Nickel detection threshold in both the {\it B} and {\it V} passbands.

Our images were reduced using a custom pipeline, as discussed by \citet{2010ApJS..190..418G}.
Host-galaxy template subtractions were performed with additional images obtained on 12 July 2016 (day 477),
after the SN had faded below the detection threshold of our telescopes.
Point-spread-function (PSF) photometry was performed with the {\tt DAOPHOT} package \citep{1987PASP...99..191S}
in the IDL Astronomy User's Library.\footnote{\url{http://idlastro.gsfc.nasa.gov/}}
Nearby reference stars in our images were calibrated to the APASS\footnote{\url{http://www.aavso.org/apass}}
catalog, which we transform to the Landolt system\footnote{\url{http://www.sdss.org/dr7/algorithms/sdssUBVRITransform.html}}
and then to the KAIT4 natural systems using the colour terms and equations as calculated by
\citet{2010ApJS..190..418G,2013MNRAS.433.2240G}.
As the Nickel camera has aged, our best-fit colour terms for the above transformation have changed;
we correct the data published here with updated Nickel colour terms recalculated in 2016
($C_B = 0.041,\ C_V = 0.082,\ C_R = 0.092,\ C_I = -0.043$).

Table~\ref{tab:photometry} presents our photometry of SN~2015G within the natural photometric systems of KAIT4/Nickel.
Because our observations show a significant gap in the bluer passbands, after the SN dropped below the
sensitivity limits for KAIT in those passbands but before we began our campaign with the larger Nickel telescope,
we do not convert these data into a standard photometric system.
\citet{2010ApJS..190..418G} and \citet{2013MNRAS.433.2240G} provide
the colour terms and equations required to perform this conversion, but doing so at all phases would require interpolating
the evolution in the bluer passbands, so we provide only the natural photometry and leave any
conversion (required for detailed comparisons with observations from other instruments) to future work. 

Infrared (IR) imaging through the $J$, $H$, and $K_s$ filters was obtained
18\,d and 35\,d after discovery with the
Wisconsin-Indiana-Yale-NOAO (WIYN) 3.5\,m telescope using the
WIYN High-Resolution Infrared Camera \citep[WHIRC;][]{2008SPIE.7014E..2WM}.
SN 2015G was clearly detected in all three bands at both epochs.
The raw images were processed using the methods described by \citet{2014ApJ...784..105W}
to construct the combined stacked images for each visit. We used Source Extractor
to obtain aperture photometry, and we calculate photometric zeropoints for these data
by cross-matching field stars with the Two Micron All Sky Survey catalog
\citep[2MASS;][]{2006AJ....131.1163S}.  We do not correct for any colour differences
between the WIYN+WHIRC and 2MASS systems.

The Ultraviolet/Optical Telescope \citep[UVOT;][]{2005SSRv..120...95R}
mounted on the {\it Swift} satellite \citep{2004ApJ...611.1005G}
was used to observe the field of SN~2015G regularly from day 12 through
day 30.  Photometric reduction for these data was performed with the pipeline for the {\it Swift} Optical Ultraviolet Supernova Archive
\citep[SOUSA;][]{2014Ap&SS.354...89B}.
For each of these images, a $5''$ aperture is used to measure the counts for the coincidence loss correction,  
and a $3''$ or $5''$ source aperture (depending on the uncertainty from above) was used for the photometry 
after subtracting off the galaxy count rate in a template image.
We apply aperture corrections (based on the average PSF in the {\it Swift} CALDB),
zeropoint corrections, and time-dependent sensitivity loss corrections to place
the magnitudes on the UVOT photometric system as described by \citet{2008MNRAS.383..627P}
and \citet{2011AIPC.1358..373B}.
Most of these observations produced nondetections of SN~2015G, which prove
useful in constraining the luminosity of SN~2015G at UV wavelengths.

The {\it HST } and its Wide Field Camera 3 (WFC3) was used to obtain optical images of
SN~2015G through the {\it F555W} band on day 20 and the
{\it F555W} and {\it F814W} bands on day 247, as part of programs GO-14149 (PI A.~Filippenko)
and GO-13683 (PI S.~Van Dyk).\footnote{Another epoch of imaging was 
attempted on 15 October 2016 as part of GO-14668 (PI A.~Filippenko), 
but unfortunately the observations were set up such that the pointing 
was toward the center of NGC~6951 and, owing to the orientation of the 
image array, the SN site itself was missed.}
We also examined pre-explosion {\it HST} observations of the SN~2015G explosion site
obtained in 2001 through the {\it F555W} and {\it F814W} filters with the
Wide-Field Planetary Camera 2 (WFPC2) as part of the campaign to monitor SN~1999el
\citep[GO-8602 with PI A.~Filippenko; see][]{2002PASP..114..403L}.
All of these images were obtained from the {\it HST} Archive after standard pipeline processing.
We performed photometry on these images using {\tt DOLPHOT} \citep{2000PASP..112.1383D}.
The {\tt DOLPHOT} parameters {\tt FitSky} and {\tt RAper} were set to 3 and 8 (respectively),
appropriate for crowded galactic environments, and we set {\tt InterpPSFlib=1},
implemented with the TinyTim PSF library \citep{2011SPIE.8127E..0JK}.

\begin{figure}
  \includegraphics[width=\columnwidth]{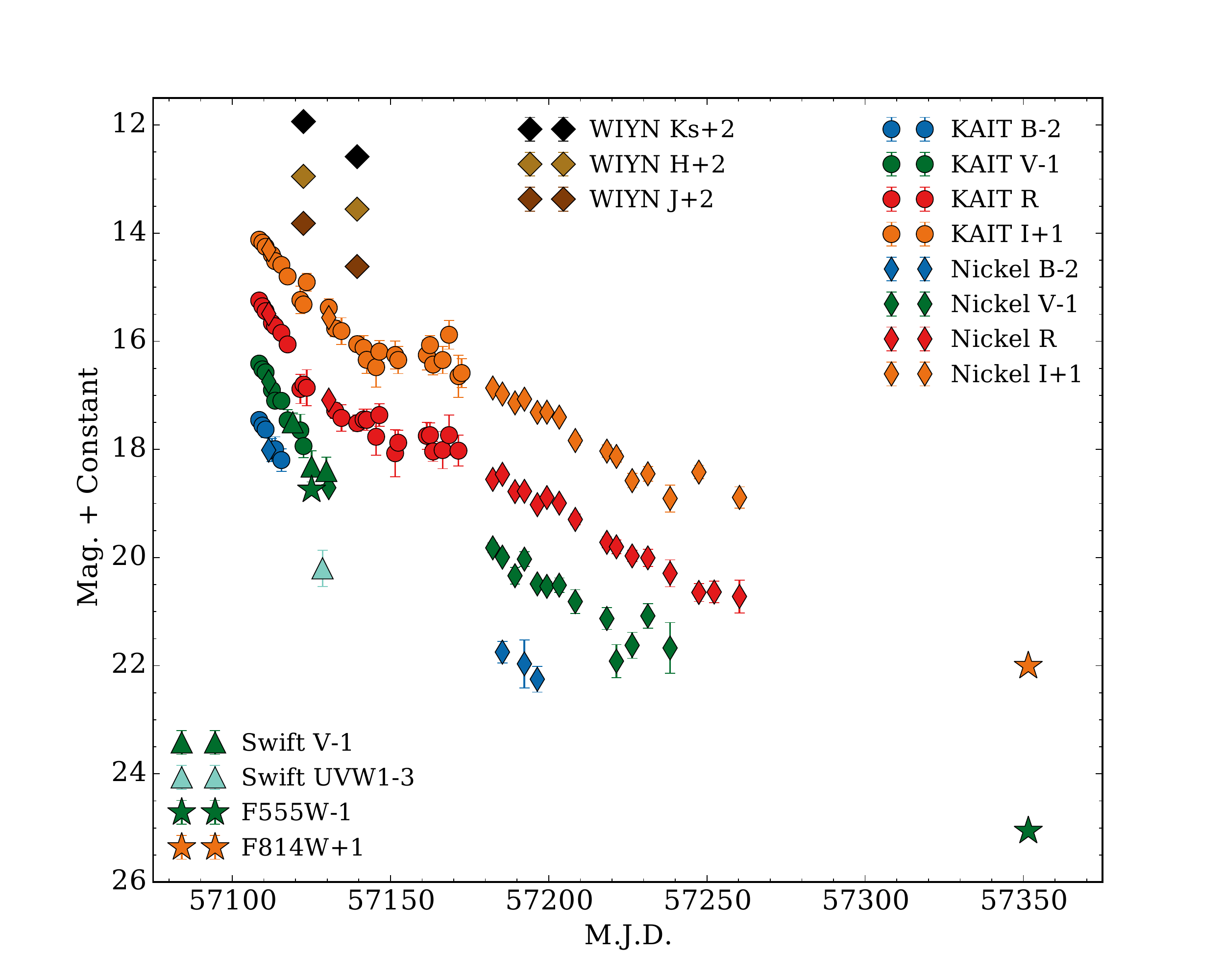}
  \caption{Our photometry of SN~2015G, from UV through IR wavelengths.
    All data are shown after correcting for extinction
    arising from dust within the Milky Way Galaxy and the SN host galaxy 
    (see \S\ref{sec:extinction}).
  \label{fig:phot} }
\end{figure}

The above data are presented in Table~\ref{tab:photometry} and shown in Figure~\ref{fig:phot}.
Values in the table are given as observed, without applying any dust reddening corrections.
Note that the error bars for our photometry are statistical only,
not accounting for systematic errors accrued (for example) during host-galaxy template 
subtraction; this may account for the larger than expected dispersion among some of the data points.
Additional UV-wavelength nondetections were obtained with {\it Swift}; we list only those
relevant for this work.

\input{phottable}

\subsection{Radio}
\label{sec:radio}

We obtained three epochs of observations on SN~2015G
using the Janksy Very Large Array (VLA), in April, May, and July 2015,
all of which were nondetections producing upper limits on the radio flux
from the SN.

All data were taken in the standard continuum-observing mode with a bandwidth of
$\rm 16 \times 64 \times 2$\,MHz.
The VLA underwent a few configuration changes at various stages during these observations.
During the data reductions, we split the
data into two side bands of approximately 1\,GHz each, centred on 4.8 and 7.1\,GHz.
We used the radio source 3C286 for flux calibration, and calibrator J2022+6136 for phase referencing.
Data were reduced using standard packages within the Astronomical Image Processing System (AIPS).
No radio emission was detected from SN~2015G in any of these observations, resulting in the deep flux limits
summarised in Table~\ref{tab:vla}.

\begin{table*}
\centering
\caption{Table of Radio Observations}
\label{tab:vla}
\begin{tabular}{cccccc}
\hline
Date		& MJD &  Phase &                 Frequency	&	$3\sigma$ RMS&	VLA 	\\
(UT)		&     &  (days) & (GHz)		& 	($\mu$Jy)		& Configuration \\
\hline
\hline
2015-04-04.42 & 57116.42& 11.64 &  4.8 	&   	$<$	36.6		&	B	\\       
2015-04-04.42 & 57116.42& 11.64 &  7.1 	&   	$<$	34.8		&	B	\\
2015-05-14.37 & 57156.37& 51.59 &  4.8 	&   	$<$	32.1		&	BnA	\\  
2015-05-14.37 & 57156.37& 51.59 &  7.1 	&   	$<$	37.8		&	BnA	\\ 
2015-07-25.19 & 57228.19& 123.41&  4.8 	&   	$<$	40.5		&	A	\\       
2015-97-25.19 & 57228.19& 123.41&  7.1 	&       $<$	29.8		&	A	\\
\hline
\end{tabular}
\end{table*}

\subsection{Ultraviolet and Optical Spectra}
\label{sec:specdata}

\begin{figure*}
    \hbox{  
      \hspace{1.5cm}
      \includegraphics[height=0.9\textheight]{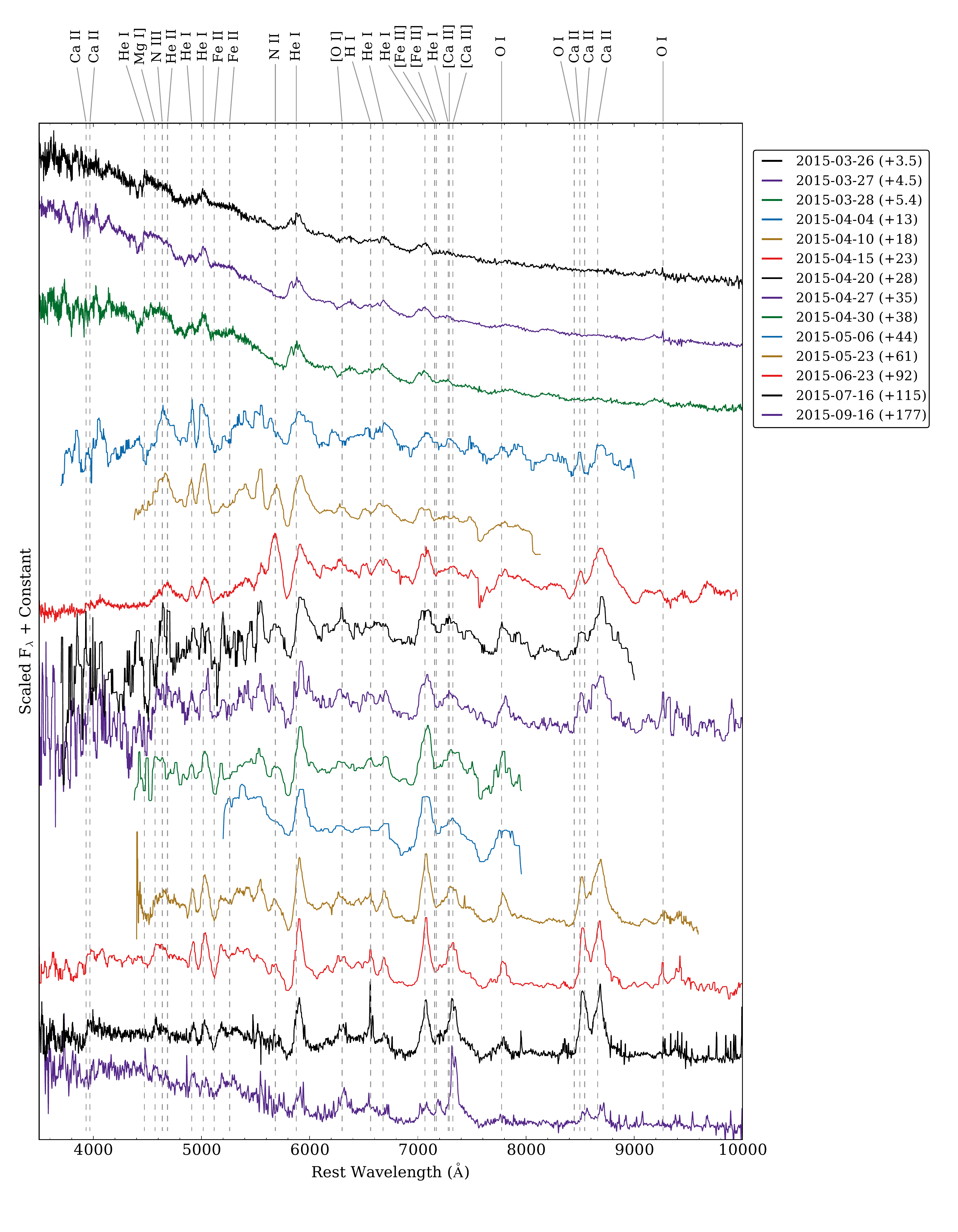}
    }
    \caption{Subset of the observed spectral series, showing only the
    higher-SNR spectra for clarity.  Our complete spectral dataset is listed in Table~\ref{tab:speclog}.
    Data have been corrected for dust absorption along the line of sight and 
    are presented in the host galaxy's rest frame, and in the legend 
    we state both the
    UT date of observation and the days since discovery.
    For some of the later epochs,
    we plot coadditions of multiple spectra to increase the SNR,
    and we label these coadditions with the mean date.
    Spectra at phases $>10$\,d have been smoothed via a moving window median or
    convolution with a Gaussian kernel. 
    All line labels are plotted at the rest wavelenth of the line ($v = 0$\,\kms).}
    \label{fig:spectra}
\end{figure*}

\input{speclog}

Regular optical spectra of SN~2015G were obtained with the
Kast Double Spectrograph mounted on the 3\,m Shane telescope \citep{kast}
at Lick Observatory, starting day 3 \citep[see][]{2015ATel.7298....1F} 
and continuing until the SN was too faint for Kast.  A single observation was
taken with Kast in the spectropolarimetric mode, on day 4; the details
of our spectropolarimetric observing techniques are described
by \citet{2015MNRAS.453.4467M} and \citet{2016MNRAS.461.3057S}.
During the same time period we obtained additional spectra with the
Boller \& Chivens Spectrograph mounted on the 2.3\,m Bok Telescope 
and the Spectrograph for the Rapid Acquisition of Transients \citep[SPRAT;][]{2014SPIE.9147E..8HP} mounted on
the 2.0\,m Liverpool Telescope.
With these telescopes, we were able to maintain a cadence between
observations of one week or less for the first two months of SN~2015G's evolution.

Several additional optical spectra  were obtained after the SN had significantly faded
using the DEep Imaging Multi-Object Spectrograph
\citep[DEIMOS;][]{2003SPIE.4841.1657F} and the
Low Resolution Imaging Spectrometer \citep[LRIS;][]{1995PASP..107..375O,2010SPIE.7735E..0RR}
mounted on the Keck 10\,m telescopes,
the Multi-Object Double Spectrograph \citep[MODS;][]{2000SPIE.4008..934B}
mounted on the 8.4\,m Large Binocular Telescope (LBT), and the
Bluechannel spectrograph on the 6.5\,m Multiple Mirror Telescope (MMT),
extending our spectroscopic sequence out until 16 September, some 6\,months
after SN~2015G was discovered.

All ground-based spectra were observed at the parallactic angle
to minimise slit losses from atmospheric refraction \citep{1982PASP...94..715F}.
We reduced and calibrated our Keck and Lick observations following the
procedures detailed by \citet{2012MNRAS.425.1789S}, utilising
IRAF\footnote{IRAF is distributed by the National Optical Astronomy
  Observatory, which is operated by AURA, Inc., under a cooperative
  agreement with the NSF.} routines and custom Python and IDL 
codes.\footnote{\url{https://github.com/ishivvers/TheKastShiv}}
For all Arizona facility telescopes (Bok, LBT, MMT), we performed standard reductions in
IRAF.  We use the standard reductions of SPRAT data as provided by the Liverpool automated pipeline.
All data were flux calibrated via spectrophotometric standards observed through an
airmass similar to that of SN~2015G, each night.  We performed the spectropolarimetric reduction
in the manner described
by \citet[and references therein]{2015MNRAS.453.4467M}, producing the reduced polarimetric parameters
of $q$ and $u$ (Stokes parameters),  $P$ (debiased polarization), and $\theta$ (sky position angle). 

As part of program GO-13797, we obtained three epochs of
{\it HST}/Space Telescope Imaging Spectrograph (STIS)
spectroscopy of SN~2015G (on 4, 11, and 20 April 2015)
covering UV through near-IR wavelengths.
We use the reduced spectra as provided by the 
Space Telescope Science Data Analysis System (STSDAS) pipeline.

Table~\ref{tab:speclog} lists our spectra, and Figure~\ref{fig:spectra} illustrates
the spectral evolution of SN~2015G.  All spectra will be made available for
download through WISeREP\footnote{\url{wiserep.weizmann.ac.il}}
\citep{2012PASP..124..668Y},
the Open Supernova Catalog\footnote{\url{sne.space}} \citep{2016arXiv160501054G},
and the UC Berkeley Supernova Database\footnote{\url{heracles.astro.berkeley.edu/sndb}}
\citep[SNDB;][]{2012MNRAS.425.1789S}.

\section{Analysis}
\label{sec:discussion}

We analyze all data in the rest frame of the host galaxy NGC 6951, adopting 
$z = 0.00475 \pm 0.000005$ \citep{1998AJ....115...62H} and a distance
of $\mu = 31.83$ (23.2\,Mpc) --- the median of 17 distances to NGC~6951 as reported in the
NASA Extragalactic Database (NED\footnote{\url{ned.ipac.caltech.edu/}}).
Note that there is a large spread in distance estimates for NGC~6951 reported in NED, ranging from 33.0 down to 16.2\,Mpc
\citep[determined via SN~Ia and Tully-Fisher methods, respectively;][]{2001A&A...372..824V,2014MNRAS.444..527S}.
Before our analysis, we correct for absorption arising from dust both within our Milky Way Galaxy and within NGC~6951.

\subsection{Line-of-Sight Extinction}
\label{sec:extinction}

\begin{figure}
  \includegraphics[width=\columnwidth]{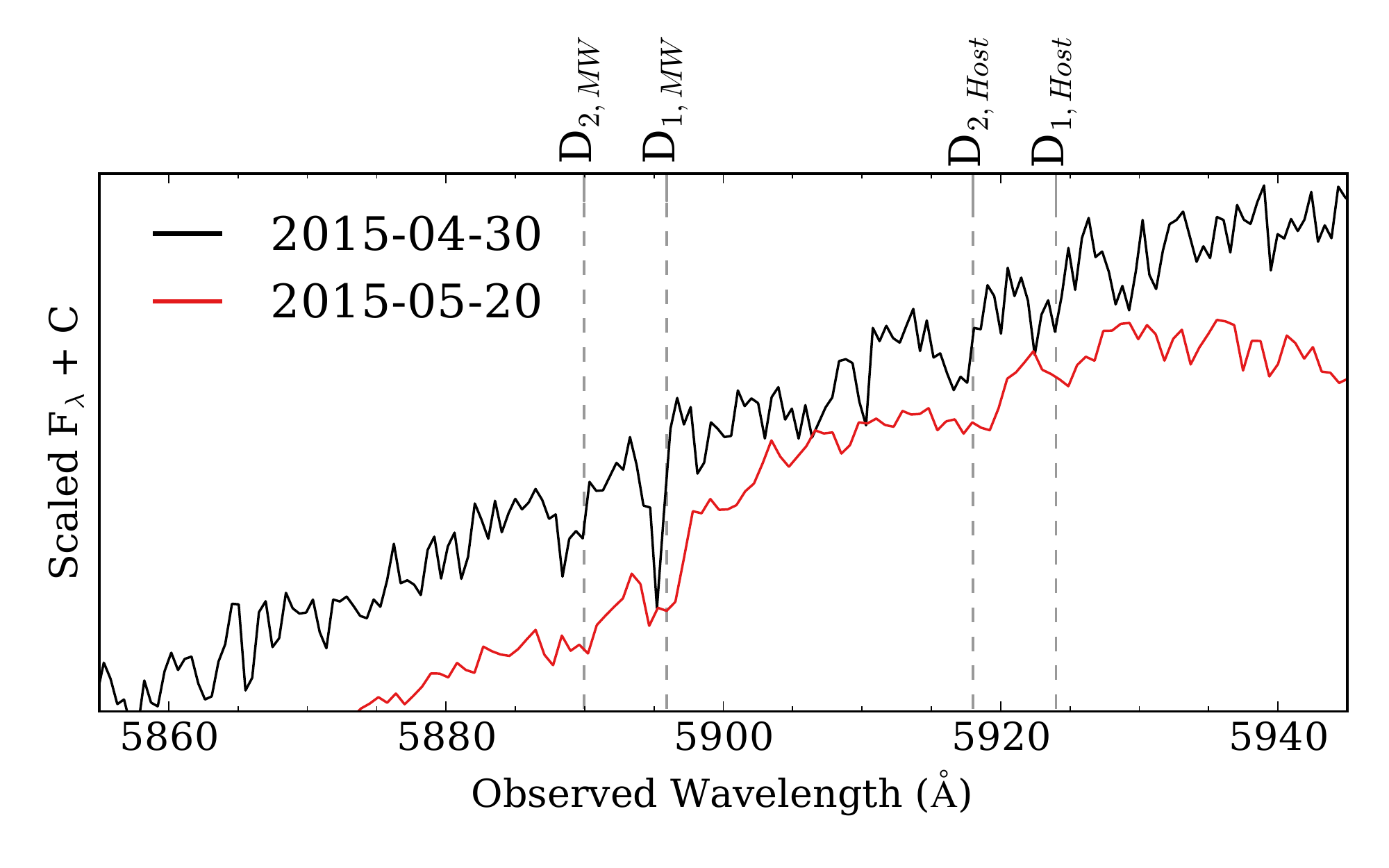}
  \caption{\ion{Na}{1}\,D absorption lines observed in two of our higher-resolution
  spectra of SN~2015G,
  with the wavelengths of the doublet indicated
  in the Milky Way rest frame and in the rest frame of the host, NGC~6951.
  \label{fig:nad} }
\end{figure}

SN~2015G lies behind a moderate amount of dust extinction arising from the
intestellar medium (ISM) within the Milky Way (MW):
$E(B-V)_{\rm{MW}} = 0.3189$\,mag \citep{2011ApJ...737..103S}.  In our spectra that exhibit sufficient 
signal-to-noise ratio (SNR) and resolution,
we observe an \ion{Na}{1}\,D absorption doublet from gas within the MW, as
well as \ion{Na}{1}\,D from gas within the host galaxy
NGC~6951.  These absorption features fall on the blue edge
of SN~2015G's \ion{He}{1}\,$\lambda$5876 emission line (see Figure~\ref{fig:nad}).

We use these sodium doublets to estimate the extinction toward SN~2015G arising within the host galaxy.
We measure the equivalent widths of the
(separately resolved) D$_1$ and D$_2$ lines in both the MMT spectrum from 30 May and the Keck spectrum from 20 June,
averaging multiple measurements from both spectra.
We obtain $0.20 \pm 0.06$ and $0.37 \pm 0.02$\,\AA\ for D$_1$ and D$_2$, respectively.
Assuming the dust and gas properties within NGC~6951 are
well approximated by their properties within the MW ($R_V = 3.1$),
we use the relations of \citet{2012MNRAS.426.1465P} to infer
$E(B-V) = 0.053 \pm 0.028$\,mag (using the D$_1$ line) and 
$E(B-V) = 0.076 \pm 0.028$\,mag (using the D$_2$ line).

Given these measures, we estimate that NGC~6951 contributes $E(B-V) \approx 0.065$\,mag, for a total line-of-sight dust
reddening of $E(B-V) \approx 0.384$\,mag. 
Our major results are not dependent upon the exact level of dust reddening, and
we caution that our calculation of the internal host galaxy's reddening is only an estimate;
the line-of-sight \ion{Na}{1}\,D and dust within the hosts of some previous SNe have been observed to be dissimilar from
those of the MW \citep[e.g.,][]{2013ApJ...779...38P,2016MNRAS.461.3057S}.

\subsection{SN 2015G's Spectral Evolution}

Figure \ref{fig:spectra} illustrates the spectral evolution of SN~2015G.
\citet{2017ApJ...836..158H} present four other spectra, providing additional coverage of the
SN evolution between days 18 and 40.
The early-time spectra of SN~2015G show the signatures observed in many young and
intermediate-age SNe~Ibn \citep[e.g.,][]{2007ApJ...657L.105F,2008MNRAS.389..113P,2017ApJ...836..158H}.
They have a strong continuum and relatively narrow
P-Cygni helium lines (absorption minima blueshifted by $\sim 1000$\,\kms)
atop broader emission.  By our third epoch of spectroscopy (+5\,d),
the broader emission lines formed a blueshifted absorption component, 
transforming into a P-Cygni line profile with absorption minima blueshifted by $\sim 8000$\,\kms.
These broader P-Cygni lines persisted throughout the
photospheric phase and into the nebular phase, at which point the continuum had faded and
the P-Cygni absorption components had disappeared, leaving behind the
emission lines of helium and calcium which dominate our spectra out to the
last observations.

\begin{figure*}
    \includegraphics[width=\textwidth]{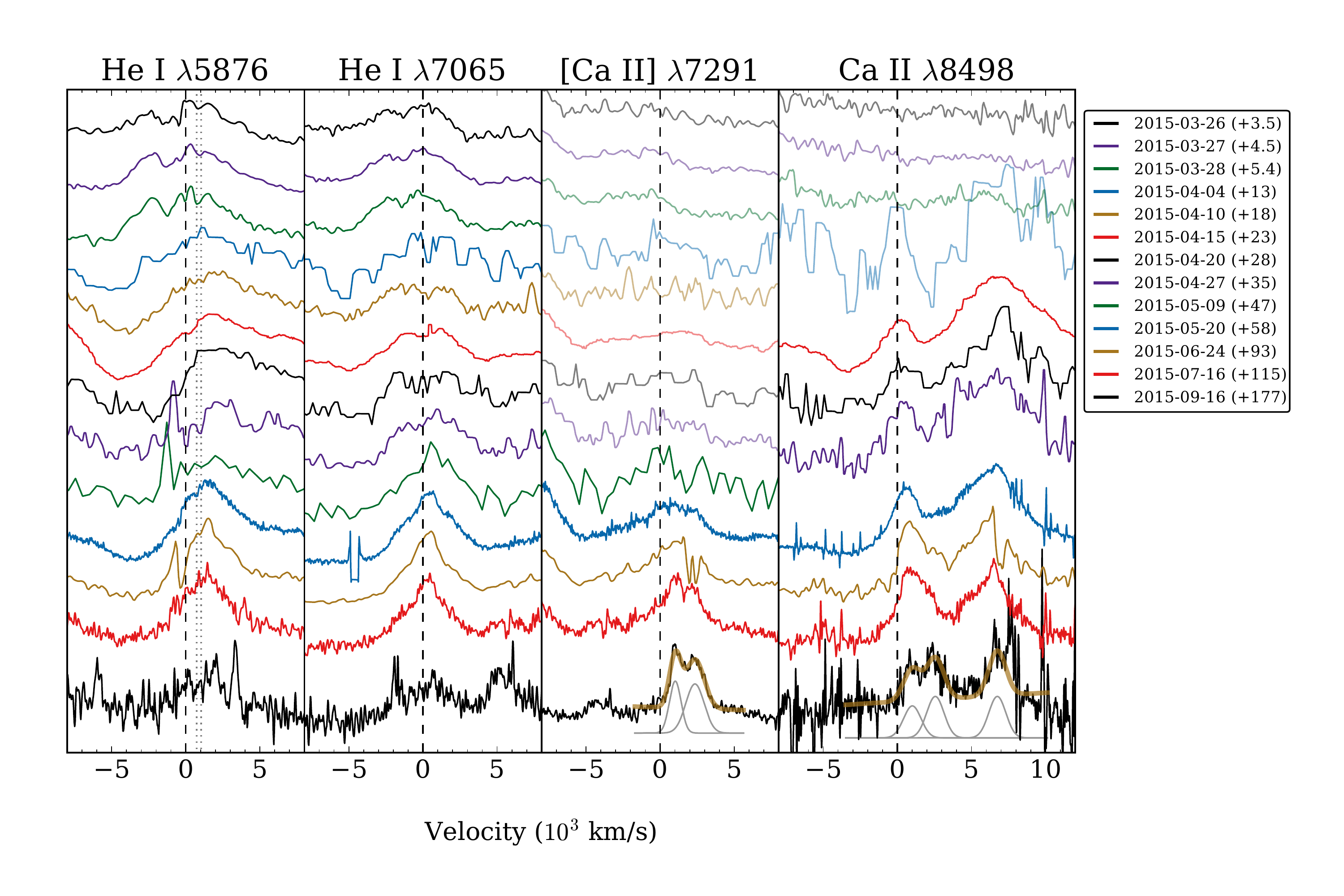}
    \caption{Evolution of \ion{He}{1}, [\ion{Ca}{2}], and \ion{Ca}{2} lines in velocity space.
    All late-time lines are strongly offset redward from 0\,\kms, most clearly the \ion{Ca}{2}. 
    Note that we calculate velocities in both calcium panels relative to the blue-most line in each blend.
    Also note that the \ion{Na}{1}\,D doublet may affect the 
    \ion{He}{1}\,$\lambda$5876 profile --- we have marked the position of this doublet in the
    left-most panel with dotted grey lines.
    Our best-fit line profiles of calcium features in the last spectrum are shown
    overlying that spectrum on the bottom, with the profile components shown below in grey.
    \label{fig:HeI} }
\end{figure*}

SN~2015G's spectral features undergo a remarkable wavelength evolution over the course
of our campaign, implying a similarly remarkable change in the velocity
of the material contributing most to those features.
Figure~\ref{fig:HeI} shows this evolution for two \ion{He}{1} lines
(including \ion{He}{1}\,$\lambda$5876, though the \ion{Na}{1}\,D
doublet may also affect this region),
for the forbidden doublet of [\ion{Ca}{2}]\,$\lambda\lambda$7291, 7324 
(which only becomes apparent starting $\sim 100$\,d, but is prominent at late phases),
and for the \ion{Ca}{2}\,$\lambda\lambda$8498, 8542, 8662 near-IR triplet.

In the \ion{He}{1} features we see early P-Cygni absorption around $-1000$\,\kms\ sitting
atop broader emission lines with widths of $\sim 5000$\,\kms.
Both the narrow and broad profiles at first
show emission components centred near velocity $v = 0$\,\kms.
By $\sim 10$\,d, the major lines at wavelengths $\lambda \lesssim 6500$\,\AA\ 
(where the continuum is strongest) show broad, blueshifted P-Cygni absorption,
as is normal in SN spectra.
By $\sim 20$\,d time the narrow P-Cygni features have faded from our spectra.

The SN then progresses toward the nebular phase and the continuum drops away below the emission
lines.  As it does so, the peaks of those emission lines clearly trend redward through our final observation.
We model the near-IR calcium triplet in our last spectrum ($+177$\,d) to measure its implied Doppler velocity.
Our model consists of three Gaussian profiles separated by the triplet's intrinsic
spacings and forced to have the same width, and we fit it to the data via 
Monte Carlo Markov Chain (MCMC)
maximum-likelihood methods.
Our best-fit profile is shown at the bottom right of Figure~\ref{fig:HeI}, and we find
a velocity offset of $\sim 1000$\,\kms\ receding (relative to the host-galaxy rest frame)
and a full width at half-maximum intensity (FWHM) of $\sim 1400$\,\kms\ (for each component line in the triplet).
The forbidden calcium doublet can also be fit quite well by a doublet profile
with an offset velocity forced to match that of the near-IR triplet, though we were required to allow
the two components of the doublet to exhibit different FWHMs;
perhaps the \ion{He}{1}\,$\lambda$7281 line and the
[\ion{O}{2}]\,$\lambda$7320 blend contribute some flux to the feature.

\subsection{Comparisons with Other SNe}

\begin{figure*}
    \includegraphics[width=\textwidth]{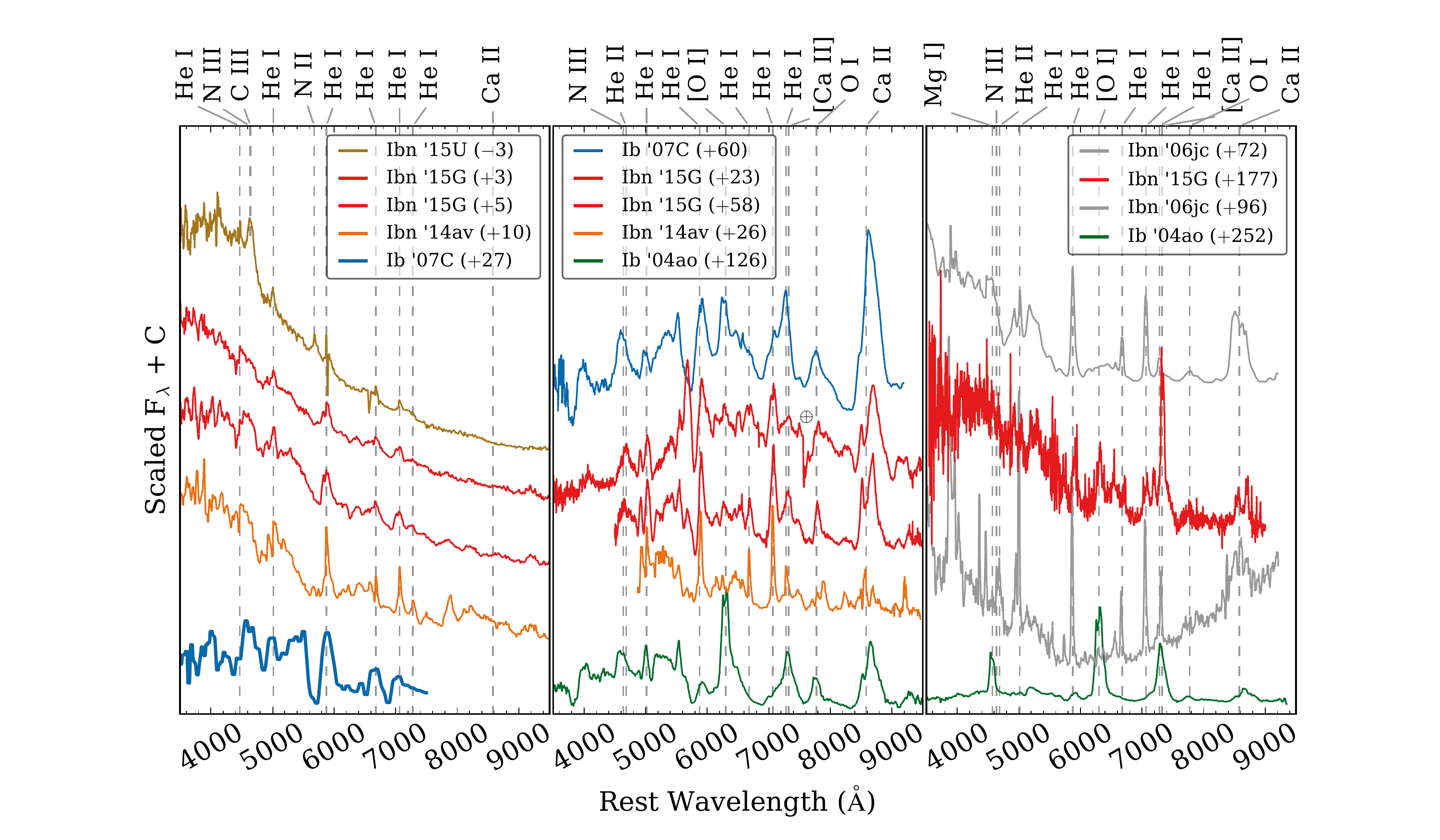}
    \caption{The spectra of SN~2015G at early, middling, and nebular phases
    (left, middle, and right panel, respectively).
    We compare to spectra of SNe~Ibn 2006jc, 2014av, and 2015U
    \citep{2007ApJ...657L.105F,2008ApJ...680..568S,2016MNRAS.456..853P,2016MNRAS.461.3057S};
    to the SN~Ib~2007C SNID template presented
    by \citet{2014arXiv1405.1437L} along with previously unpublished spectra of the event,
    and to previously unpublished spectra of SN~Ib~2004ao.
    For each SN, spectra are presented in the SN rest frame and have
    been corrected for dust reddening along the line of sight.  The phases of the
    spectra are given in days after discovery for SNe~2004ao, 2006jc, and 2015G,
    and days after optical peak for SNe~2007C, 2014av, and 2015U.
    All line labels are plotted at the rest wavelength of the line ($v = 0$\,\kms).
    \label{fig:spec_comparisons} }
\end{figure*}

Figure~\ref{fig:spec_comparisons} shows spectra of SN~2015G compared to those
of a few other SNe~Ibn and normal SNe~Ib, with early-time spectra plotted on the left, 
late-time spectra on the right, and those taken at intermediate ages in the middle.
We ran our +5\,d spectrum of SN~2015G through the SuperNova Identification code \citep[SNID;][]{2007ApJ...666.1024B}
with the updated template sets of \citet{2012MNRAS.425.1789S} and \citet{2014arXiv1405.1437L}.
Note that none of these template sets includes SN~Ibn spectra, but 
SNID identified the +27\,d SN~Ib~2007C spectrum as a reasonable match.
Broad features arising from many of the same ions are apparent in both SNe, though 
of course the SN~2007C spectrum
(which is reasonably characteristic of normal SN~Ib spectra at photospheric phases)
does not exhibit the strong blue continuum and narrow emission features of SNe~Ibn.
The presence of a similar set of broad features between these two SNe suggests that SN~2015G 
could have been a relatively normal SN~Ib if its dense CSM were not present.

The pre-maximum spectrum of SN~2015U showed emission lines of
\ion{N}{2}\,$\lambda\lambda$5680, 5686 and \ion{N}{3}\,$\lambda\lambda$4634, 4642/\ion{C}{3}\,$\lambda$4647
\citep{2015MNRAS.454.4293P,2016MNRAS.461.3057S}.
Neither of these is apparent in our early spectra of SN~2015G. 
However, probable \ion{N}{2}\,$\lambda\lambda$5680, 5686 and 
\ion{N}{3}\,$\lambda\lambda$4634, 4642 emission features do become
prominent by day 23 (see Figure~\ref{fig:spectra}).

During its transition from the photospheric phase to the nebular
(see the middle panel of Figure~\ref{fig:spec_comparisons}), SN~2015G followed
an evolution not so dissimilar from that of normal SNe~Ib, as the previously dominant continuum faded
away to leave only the nebular emission lines behind.
However, the ``broad'' nebular features of SN~2015G are less broad than
those shown by normal SNe~Ib throughout, and the strong \ion{He}{1} emission
lines of SN~2015G do not appear in normal SN~Ib spectra.
A similar spectral evolution, starting with a strong blue continuum topped by
narrow features and ending with the nebular emission features 
commonly observed in normal SNe~Ib, has been seen in several SNe~Ibn
\citep[e.g., SN~2010al and ASASSN-15ed, both of which share many similarities with SN~2015G;][]{2015MNRAS.449.1921P,2015MNRAS.453.3649P}.

This strong \ion{He}{1} emission is usually interpreted as a result
of ongoing interaction between the outer layers of the ejecta and
a helium-rich CSM \citep[e.g.,][]{2000AJ....119.2303M}.
Because the excitation energy of \ion{He}{1} is so large,
thermal excitation in the freely expanding and cooling SN ejecta
cannot account for features like these.
Instead, the ongoing shock between the ejecta and the helium-rich CSM
is usually invoked to explain the strong helium lines in SNe~Ibn.
Normal SNe~Ib, which show strong \ion{He}{1} absorption lines at peak
but no nebular \ion{He}{1} emission lines, are often argued to arise via
a different process: ``mixed'' SNe \citep[e.g.,][]{1991ApJ...383..308L}.
In this scenario, radioactive nickel is mixed well out into the helium envelope
and the $\gamma$ rays from the decay process are therefore able
to excite the helium nonthermally.

\citet{2017ApJ...836..158H} propose the existence of two spectroscopic
subclasses of SNe Ibn: those that show narrow ($v \approx 1000$\,\kms) \ion{He}{1}
P-Cygni features within $\sim 20$\,d of peak and those that do not.
They group SNe~2015G and 2015U into the first class and SN~2006jc into the second.
SN~2014av (also shown in Figure~\ref{fig:spec_comparisons}) they label as
uncertain.
The spectra presented here show that the similarities between SNe~2015G and 2006jc
are manyfold and robust while also confirming that SN~2015G did show narrow P-Cygni \ion{He}{1}
lines at early phases.

This adds evidence to the argument that either SNe Ibn exhibit a continuum of spectral properties
between the two extremes of the \citet{2017ApJ...836..158H} subclasses,
or that the issue of interest is not the presence or absence of narrow P-Cygni
lines but instead their longevity or our viewing angle.
Perhaps all SNe~Ibn show these features but they quickly disappear from the spectra of
SN~2006jc-like events, as those authors speculate, or perhaps SNe Ibn often
exhibit strong asymmetries (in their CSM distributions or in their explosions)
and the narrow lines are only apparent from certain perspectives.

SN~2015G exhibited much stronger calcium (both allowed and forbidden) than did
SN~2014av, though comparisons to SN~2006jc indicate that the strength of these calcium features
are quite variable among different SNe~Ibn.
As the SNe~Ibn shown in Figure~\ref{fig:spec_comparisons}
enter the nebular phase, their [\ion{O}{1}]\,$\lambda\lambda$6300, 6364
emission lines are weak compared to those of normal SNe Ib.
The [\ion{O}{1}] line strengths of SN~2015G are between those
of the very oxygen-weak SN~2006jc and the oxygen-strong (normal)
Type Ib SN~2004ao, at both mid-nebular and fully-nebular phases.

As SN~2015G aged a blue continuum became apparent,
similar to those observed in the Type Ibn SNe~1999cq and 2006jc
\citep{2000AJ....119.2303M,2007ApJ...657L.105F} and the transitional
IIn/Ibn SN~2011hw \citep{2012MNRAS.426.1905S},
and visible in the right panel of Figure~\ref{fig:spec_comparisons}.
The late-time blue continuua of SNe~1999cq and 2006jc were
attributed to a forest of blended \ion{Fe}{2} lines;
we believe the same process to be at work in the SN~2015G system.
The density of features at these wavelengths makes individual lines
difficult to isolate, but we identify clear \ion{Fe}{2} features by comparing our 15 April spectrum to
synthetic models calculated with SYN++ \citep{2000PhDT.........6F,2011PASP..123..237T}.
We are unable to converge upon a SYN++ model that reproduces all of the major spectral features,
most especially the strong \ion{He}{1} emission, but 
our best-fit model does argue for significant near-UV line blanketing from iron as well
as multiple overlapping absorption features in the blue, the two most
obvious and isolated of which are indicated in Figure~\ref{fig:spectra}.

This blue pseudocontinuum
arose later in SN~2015G's evolution than it did for SN~2006jc.
Though the peak dates of both SNe passed by unobserved, the relative strengths of the 
\ion{Ca}{2} and [\ion{Ca}{2}] lines, as well as the \ion{O}{1} and [\ion{O}{1}]
features, provide another indicator of the age.
As the SN aged, forbidden emission lines became prominent while those lines
arising from allowed transitions faded.
The right-most panel of Figure~\ref{fig:spec_comparisons} shows that the
blue pseudocontinuum of SN~2006jc formed while the near-IR triplet of \ion{Ca}{2}
was strong, but the blue pseudocontinuum of SN~2015G did not become
apparent until the near-IR triplet had faded and [\ion{Ca}{2}] emission had become
dominant (see also Figure~\ref{fig:spectra}).  
This enhanced blue continuum is generally understood to arise via
nonthermal excitation of iron via an ongoing shock as ejecta
collide with CSM \citep[e.g.,][]{2007ApJ...657L.105F}, and
normal SNe Ib do not show it.

SN~2006jc, at extremely late phases, developed a red/near-IR continuum as
dust formed in the shocked shell of material created
by the SN ejecta's collision with the CSM
\citep{2008ApJ...680..568S,2009ApJ...692..546S}.
This red continuum is apparent 
in the $+$96\,d spectrum of SN~2006jc; our spectra of SN~2015G show that no
such late-time red continuum formed in SN~2015G during our 6 months of
spectroscopic follow-up observations.

Interestingly, \citet{2008ApJ...680..568S} identified transient
\ion{He}{2}\,$\lambda$4686 and \ion{N}{3}\,$\lambda\lambda$4634, 4642 emission
with the rise of the red hot dust continuum of SN~2006jc, tying these phenomena
to observations of the colliding winds in $\eta$ Carinae \citep{2010MNRAS.402..145S}.
The middle and right panels of Figure~\ref{fig:spec_comparisons} show that
the same transient emission feature of ionized helium and nitrogen was temporarily present
in SN~2015G but at a much earlier phase, and it does not appear to be associated
with the formation of dust in this event.
Perhaps dust was able to form behind the shock in the SN~2006jc system,
but not in SN~2015G, because the interaction occured in SN~2006jc only after
the equilibrium temperature had dropped substantially.  
The spectra of SN~2015G showing \ion{He}{2} and \ion{N}{3} emission still
exhibit a best-fit blackbody temperature of $\sim 4000$\,K --- likely too hot
to allow any dust formation in the system at that phase.

\begin{figure}
    \includegraphics[width=\columnwidth]{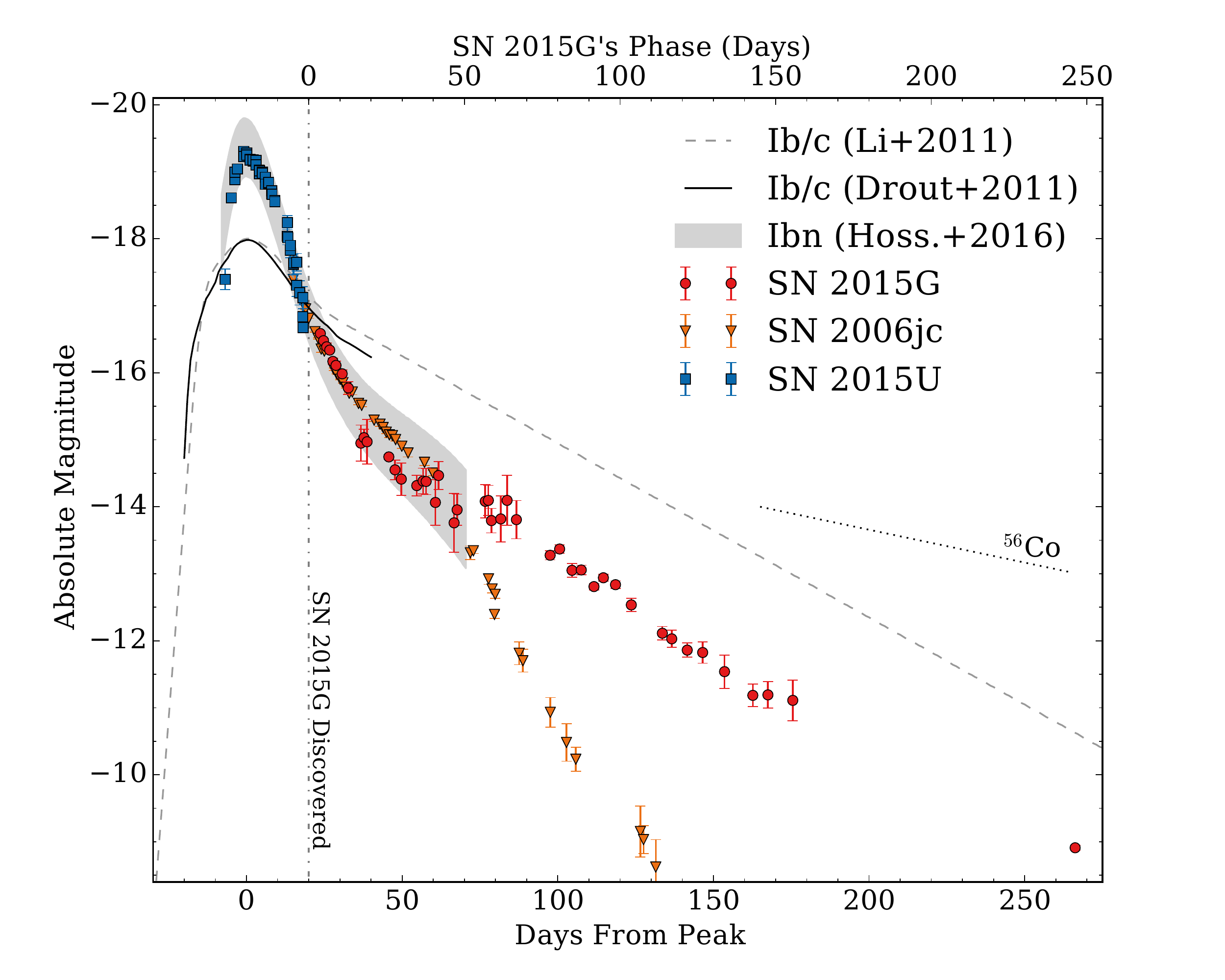}
    \caption{ Absolute {\it R}-band light curve of SN~2015G, compared
    with absolute {\it R} light curves of
    SNe~Ibn 2006jc \citep{2007ApJ...657L.105F,2016ApJ...833..128M}
    and 2015U \citep{2016MNRAS.461.3057S}.  We compare to the template light curves
    of \citet[dashed line]{2011MNRAS.412.1441L},  \citet[solid line]{2011ApJ...741...97D},
    and \citet[shaded region]{2017ApJ...836..158H}.
    We offset the SNe 2006jc and 2015G light curves so as to best
    match with the SN~Ibn template, and we show the phase of SN~2015G
    relative to its discovery date on the top axis with the inferred
    phase relative to optical peak on the bottom axis.  We also show the 
    $^{56}$Co $\rightarrow$ $^{56}$Fe decline rate for comparison.
    \label{fig:lc_comparisons} }
\end{figure}

Figure~\ref{fig:lc_comparisons} shows the light curve of SN~2015G compared to
template light curves for SNe Ib/c and Ibn
\citep{2011MNRAS.412.1441L,2011ApJ...741...97D,2017ApJ...836..158H}
and to the light curves of SNe Ibn 2006jc and 2015U \citep{2007ApJ...657L.105F,2016MNRAS.461.3057S}.
The date of maximum brightness is unknown for both SNe~2006jc and 2015G;
for this comparison we infer the dates of maximum for these two SNe
by comparing their observed light curves to the template of \citet{2017ApJ...836..158H}
while respecting the limits imposed by what constraints we have on their explosion dates.
We estimate that SN~2006jc was discovered some 6\,d after peak and SN~2015G
was discovered 20\,d after peak.

The photometric evolution of SN~2015G is similar to that of other SNe~Ibn
\citep[e.g.,][]{2016MNRAS.456..853P,2017ApJ...836..158H}.
The optical light curves decline roughly
linearly and slightly more rapidly than the $^{56}$Co $\rightarrow$ $^{56}$Fe decline rate, 
as do those of normal SNe~Ib.  There is some indication of a
flattening in the {\it R} and {\it I} passbands around MJD 57170
($\sim70$ days after discovery), most clearly shown in Figure~\ref{fig:phot}.
Unfortunately, SN~2015G was, at this time, fading below our KAIT detection threshold, and our monitoring campaign using larger telescopes had not yet begun;
the data from these epochs are very noisy, and it is difficult to determine whether this
apparent flattening is genuine. If real, it could be attributed to a short-lived luminosity
enhancement from ongoing CSM interaction, but our continuing light curves do
not show any evidence for similar episodes at later times.

Note, in Figure~\ref{fig:lc_comparisons}, the steepening of SN~2006jc's {\it R}-band light curve around 75\,d,
likely caused by obscuration from a newly formed dust shell \citep{2008ApJ...680..568S}.
Observations of the SN~Ibn ASASSN-15ed suggest that it also became obscured
by a forming dust shell $\sim 2$\,months after peak \citep{2015MNRAS.453.3649P}.
In good agreement with our spectra, the light curve of SN~2015G shows no evidence for 
a forming dust shell at least out to $\sim 8$\,months after discovery.
Few SNe~Ibn have been monitored out to such late phases, so it is difficult
to know how common the formation of a dust shell may be, but, like SN~2015G,
SNe~Ibn tend to be subluminous at late phases compared to normal stripped-envelope
SNe \citep{2017ApJ...836..158H}.
If the luminosity of SNe~Ibn on the post-peak decline is driven by radioactive decay, as is the case for normal
stripped-envelope SNe, this implies that a significantly smaller amount of $^{56}$Ni 
is present in the ejecta of SNe~Ibn than in those of normal SNe~Ib/Ic.

\subsection{Spectropolarimetry}

\begin{figure}
    \includegraphics[width=\columnwidth]{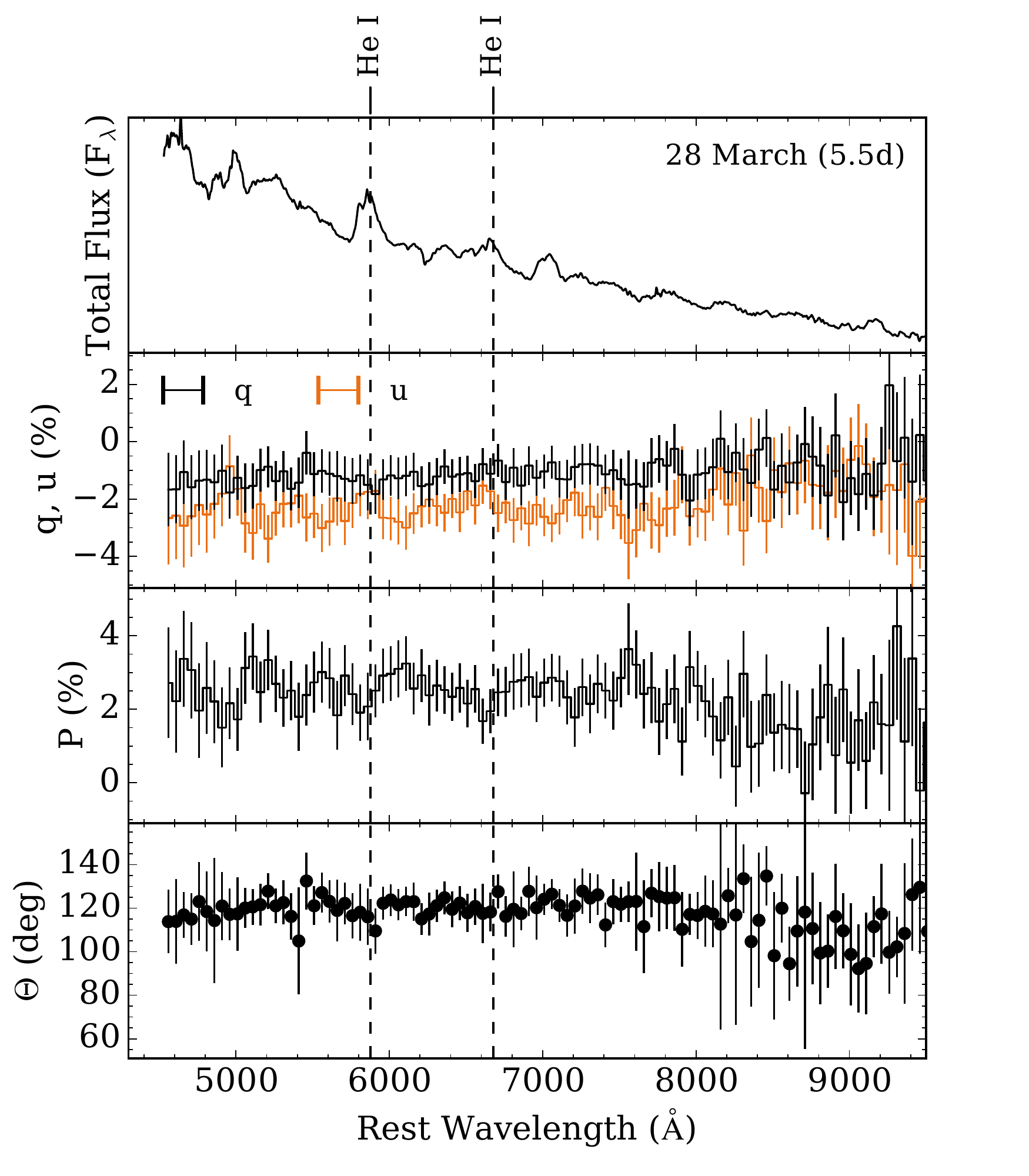}
    \caption{A single epoch of spectropolarimetry obtained 5.5\,d after the discovery of SN~2015G.
    The top panel illustrates the total-flux spectrum smoothed with a 50\,\AA\ Gaussian kernel, while
    the middle panels show the Stokes parameters ($q, u$) and total polarization ($P$) binned to 50\,\AA,
    and the bottom panel gives the measured position angle, also binned to 50\,\AA.
    \label{fig:specpol} }
\end{figure}

Our spectropolarimetric observation of the young SN~2015G is shown in Figure~\ref{fig:specpol}.
SN~2015G shows strong continuum polarization ($P \approx 2.7$\%), with some small but apparently
significant variations across the strong \ion{He}{1} line features. 

It is possible that the continuum polarization level is substantially affected or perhaps even dominated
by interstellar polarization (ISP) along the line of sight. The host galaxy's contribution
to interstellar dust absorption is relatively low and the SN appears to be in a rather remote
environment, so we do not expect the host ISP to be dominant. Nonetheless, the maximum polarization
from the host could be up to 0.6\% \citep[following][]{1975ApJ...196..261S}, with an entirely
unknown position angle. The larger MW dust absorption toward SN~2015G indicates that our
Galaxy's ISP contribution could be as high as 2.9\% \citep{1975ApJ...196..261S}.
Indeed, SN~2015G is at a low Galactic latitude of 14.8$^{\circ}$, and substantial
Galactic cirrus is present in this region of the sky. 

The MW ISP can be estimated by measuring the polarization of stars that
(1) are along the line of sight toward the SN (ideally within 1$\degree$);
(2) are suspected to have negligible intrinsic polarization
\citep[ideally spectral types A5 through F5;][]{2002PASP..114.1333L}; and
(3) lie at sufficient distances such that all ISM along the line of sight and
within a scale height of 150\,pc above the Galactic disk is sampled.
For the line of sight near SN~2015G this minimum required probe distance is 675\,pc.
Unfortunately, there are no stars in the literature of known spectral type with
polarization measurements satisfying all of these criteria. Loosening these constraints
to allow stars of any spectral types within $5\degree$,
we identify two stars in the catalog of \citet{2000AJ....119..923H} --- HD\,197911 (B5\,V; 1043\,pc)
and HD\,198781 (B0.5\,V; 712\,pc) --- which exhibit an average ($1\sigma$) polarization and
position angle of 1.36\%\,(0.03\%) and 150$\degree$\,(16$\degree$).
This position angle value is consistent with that measured for SN~2015G, while the level of polarization
is about half that of the SN. However, we are reluctant to trust these values as accurrate measures
of the ISP because at least one of those stars (HD\,197911) has been associated with a dusty
interstellar bow shock that is likely to scatter the star's light and exhibit its own polarization
\citep{2015A&A...578A..45P}.
The existence of a star at that distance (well beyond the 675\,pc limit we imposed above)
with shocked ISM in its vicinity suggests that the scattering and polarizing effect of the ISM
could extend to distances larger than expected in this region of the sky, and may be
highly spatially variable. 

To improve our census of the ISP, we obtained new Lick/Kast observations (on 2017 March 3) of two
additional probe stars of known spectral types that have smaller angular separations, $<2^{\circ}$:
BD\,661309 (A5\,V) exhibits $P_V=0.22\,(0.01)$\% and ${\theta}_V=54.7\degree\,(1.0\degree)$,
and HD\,197344 (B8\,V) exhibits $P_V=0.22\,(0.01)$\% and ${\theta}_V=64.6\degree\,(1.2\degree)$.
These stars have spectroscopic parallaxes that indicate distances of 525 and 575\,pc, respectively
--- close to, but slightly below, the minimum suggested distance to effectively probe the bulk
of intervening ISM. Nonetheless, the measured values are very low, which indicates that either the
ISP is small near the SN's line of sight or that there is substantial ISP originating from
greater distances than we are currently able to probe.

In conclusion, the complexity of the ISM in this region of the sky and the lack of excellent probe stars
has proven problematic for our efforts to obtain a reliable estimate of the ISP toward SN~2015G.
However, we note that none of the ISP estimates we have considered come close in strength to the
very strong $\sim 2.7$\% measured for the SN. It appears that, if the SN is not intrinsically polarized,
then the ISP vector components of the MW and the host must be constructively interfering
(i.e., have similar position angles, or at least be in similar quadrant of the $q-u$ plane)
to give us such a strong polarization measure. For this reason, it seems plausible that the
intrinsic polarization of the SN is significant and, therefore, that the electron-scattering
photosphere of the explosion is substantially aspherical, consistent with the other proxies
for explosion asymmetry we consider in this paper. Without better constraints on the ISP,
however, it is difficult to quantify the degree of asphericity.

\subsection{The Ultraviolet Spectra}

\begin{figure*}
    \includegraphics[width=.9\textwidth]{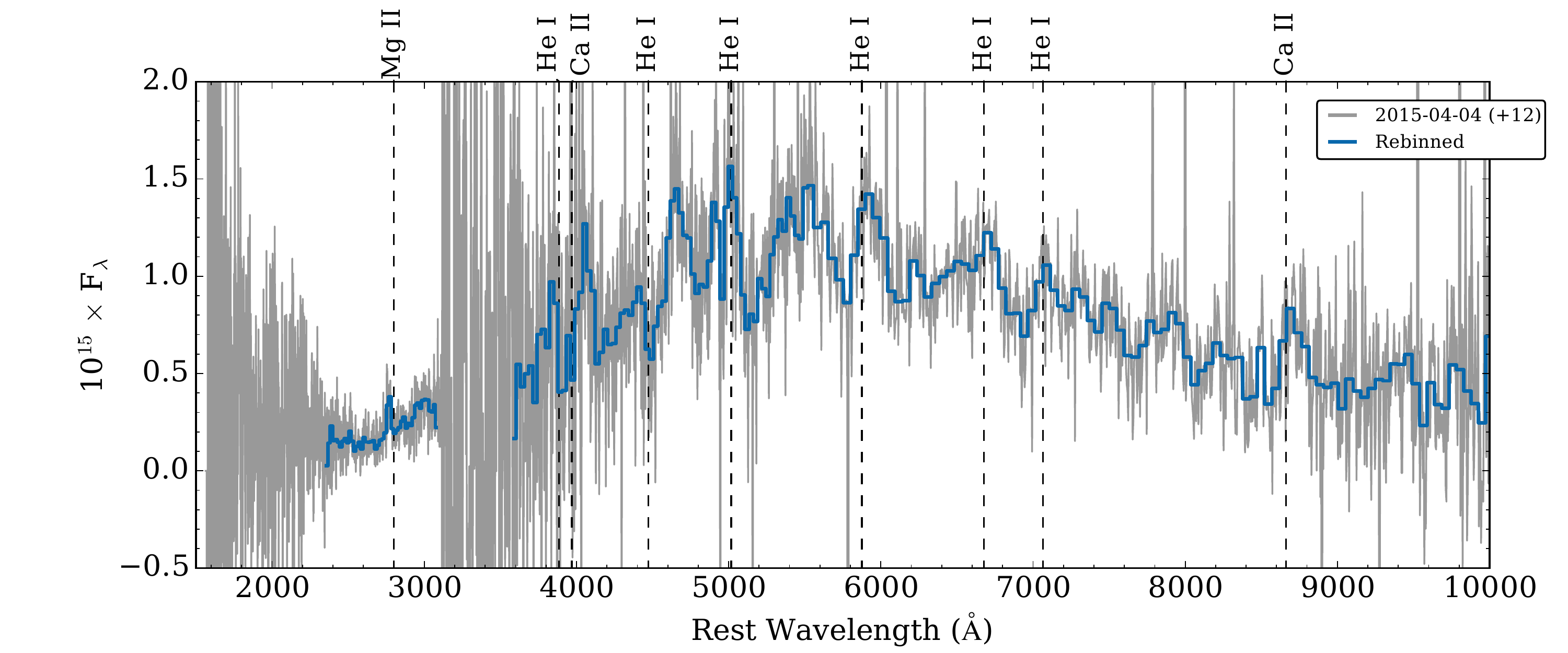}
    \caption{The UV through near-IR spectrum of SN 2015G, as observed by {\it HST}.
    Data have been corrected for dust absorption along the line of sight and 
    are presented in the host galaxy's rest frame. The full spectrum is shown in grey
    and a rebinned spectrum with bins $\sim 50$\,\AA\ wide is shown in blue.
    \label{fig:uvspec} }
\end{figure*}

Figure~\ref{fig:uvspec} shows the full STIS UV+optical
spectrum from day 13, rebinned and median-averaged in
wavelength bins $\sim 50$\,\AA\ wide to increase the SNR and reduce
the effects of cosmic rays.
Continuum emission is detected from $\sim 2300$\,\AA\ out to $\sim 1$\,${\rm \upmu}$m,
overlain by the broad and narrow P-Cygni features described above.
We find one emission line in the near-UV, labeled
\ion{Mg}{2}\,$\lambda\lambda$2796, 2803 in Figure~\ref{fig:uvspec}.
This feature is also observed in our spectrum from 11 April, but
not in the spectrum from 20 April (at which point the continuum
has faded below detectability at these wavelengths as well).
Between the first two spectra it evolves from a wavelength of
$2764 \pm 2$\,\AA\ to $2784 \pm 6$\,\AA\ 
(uncertainty estimated via MCMC fits of two Gaussian profiles separated
by the spacing of this doublet, 7.16\,\AA).
Assuming our line identification is correct, and that both lines of the doublet contribute equally to the
line flux, this implies velocity blueshifts for this feature
of about 3800\,\kms\ and 1600\,\kms\ on 4 and 11 April (respectively), so
the slowdown is $\sim 300$\,\kms\,d$^{-1}$.
No other narrow emission lines in our dataset on SN~2015G show this sort 
of behaviour and we note that it is peculiar. However, the SNR is low in our UV spectra
and, though inspection of the raw two-dimensional frames shows them to be clean with no obvious artifacts,
we are hesitant to infer too much from this putative \ion{Mg}{2} line.

\subsection{The Spectral Energy Distribution}
\label{sec:sed}

\begin{figure}
    \includegraphics[width=\columnwidth]{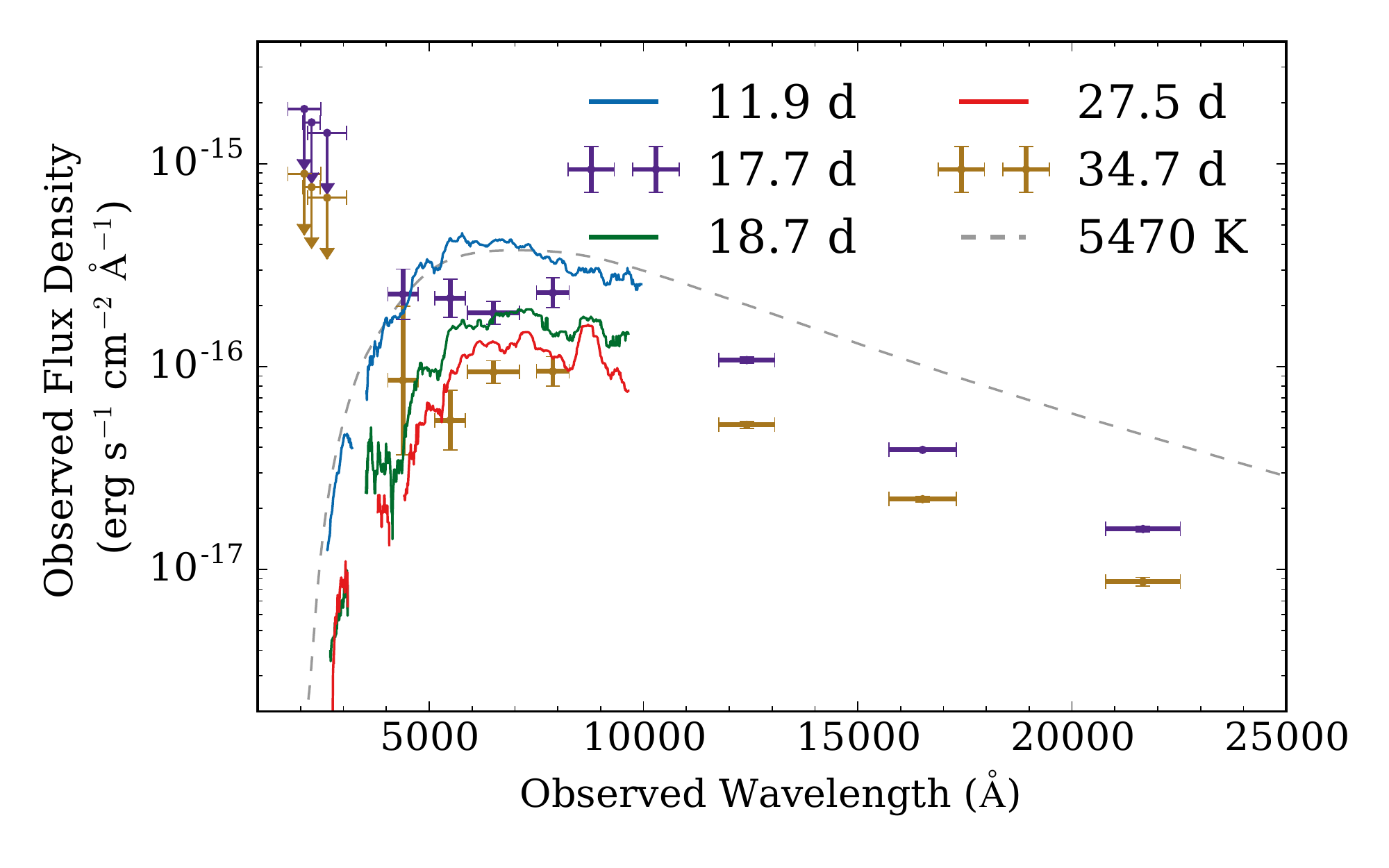}
    \caption{UV through IR observations of SN~2015G, 
    including three UV-optical spectra observed by {\it HST}, which have
    been trimmed and
    smoothed with a wide ($\sim 500$\,\AA) median box filter to illustrate the
    overall SED, alongside two epochs of
    UV--IR photometry. 
    All data are shown in the observer frame,
    with no corrections for extinction applied.
    For comparison, a blackbody at 5470\,K is
    shown in dashed grey, redshifted into the observer's frame and obscured
    by the dust populations described in \S\ref{sec:extinction}.
    Phase (days) or $T_{\rm BB}$ (K) are given in the legend.
    \label{fig:SED} }
\end{figure}

Our observations of SN~2015G cover radio through UV wavelengths,
and we are able to reconstruct a broad-wavelength spectral 
energy distribution (SED) at several phases. 
Figure~\ref{fig:SED} shows the UV through IR SED of SN~2015G as observed at 5 epochs
between 12\,d and 35\,d after discovery.
For the two phases at which we have IR photometry, we interpolate the optical
light curves to the time of the IR observations using a Gaussian process regression.
At the earlier epoch, the {\it Swift} satellite observed the location of SN~2015G
nearly concurrently (within a few hours),
and we plot the $3\sigma$ upper limits from the resulting UV nondetections.
For the later epoch we plot the last UV nondetection from {\it Swift}, which 
was observed 5\,d before the listed phase (when the optical+IR images were taken).

Our {\it HST} spectra strongly constrain the wavelength of peak flux, and we
show a best-fit model blackbody spectrum in Figure~\ref{fig:SED}, for comparison.
Fitting a redshifted and dust-reddened blackbody
(assuming the dust properties presented in \S\ref{sec:extinction})
to the first {\it HST} spectrum, we find that the SED of
SN~2015G is approximated quite well by a single-component blackbody with
a temperature of $T_{\rm BB} \approx 5470 \pm 250$\,K.
Our uncertainties about the dust reddening arising within the
host galaxy likely dominate our temperature errors, so we estimate 
the above error bars on $T_{\rm BB}$ by ranging $E(B-V)_{\rm host}$ from 0.0\,mag to 
twice our best-guess value of 0.065\,mag and refitting.
For Figure~\ref{fig:SED} we have converted our photometric observations into
flux units using PySynphot and the published filter curves for each instrument,
assuming a 5470\,K blackbody spectrum.

Between 12\,d and 35\,d after discovery,
the SED qualitatively behaves like a cooling blackbody,
fading in both temperature and luminosity. 
Our IR photometry argues for a steeper Rayleigh-Jeans tail than
do our fits to the UV/optical peak, but (given our uncertainties about the
degree and wavelength dependence of the dust obscuration toward 
SN~2015G) we are hesitant to assign much significance to that discrepancy.
We estimate the bolometric energy output of SN~2015G
and the implied blackbody radius at 11.9\,d after
discovery, based upon our blackbody fit to the {\it HST} spectrum:
$L \approx 10^{42}$\,erg\,s$^{-1}$ and $R \approx 10^{15}$\,cm.


\subsection{Limits on Radio Luminosity}

Radio emission from SNe predominantly arises via the synchrotron mechanism
as the forward shock ploughs through the CSM.
The narrow emission features in our early-time optical spectra provide
clear evidence for a dense CSM near the progenitor at the time of core collapse.
However, the density profile of the CSM at larger radii is quite
uncertain, so the radio flux expected from the SN at intermediate and late phases
is similarly uncertain.
Though our attempts to observe radio emission SN~2015G yielded 
only upper limits, they do
provide some interesting constraints on the extended CSM surrounding the SN.

We argue elsewhere in this paper that the dense CSM that made SN~2015G
a SN~Ibn was likely not created by wind-like mass loss from the progenitor, but rather 
was built up through one or more extreme mass-loss events $\lesssim$1\,yr before core collapse.
However, we find it plausible (in the absence of evidence to the contrary) to assume a history of more stable wind-like mass
loss from the progenitor at earlier times before core collapse, as is normal for the progenitors of stripped-envelope SNe
\citep[e.g.,][]{1998ApJ...499..810C,2014ApJ...797....2K,2016ApJ...818..111K}.

Our mid-phase spectra show broad lines
with absorption edges falling at blueshifts of $\sim 8000$\,\kms\ (see Figure~\ref{fig:HeI}),
placing a lower limit on the velocity of the forward shock ($v_{\rm shock} \gtrsim 8000$\,\kms),
while the narrow P-Cygni features in our early-time spectra have characteristic velocities of
1000\,\kms.  We construct a simple model of the SN~2015G system by assuming 
a history of mass loss with $v_{wind} \approx 1000$\,\kms\ and adopting $v_{\rm shock} = 10,000$\,\kms.
We further estimate that the SN took $\sim 5$\,d to rise to maximum brightness, yielding a best-guess
explosion date of 2015-02-27 (MJD 57080).
SN~2015G's rise time is effectively unconstrained; we motivate our choice by noting
that other SNe~Ibn for which the rise has been observed exhibit values of $\sim 5$\,d \citep{2016MNRAS.456..853P,2017ApJ...836..158H}.
This value is, however, quite uncertain, and SN~2015G may have taken significantly longer
to rise to peak \citep[see, e.g., the 16\,d rise shown by SN~2010al;][]{2015MNRAS.449.1921P}.

Our naive blackbody model from \S\ref{sec:sed} showed that our data
are described rather well by a cooling blackbody of $R \approx 10^{15}$\,cm.  
If we take this value to be the outer extent of the low-radii dense CSM, the forward shock
would have taken $\sim 10$\,d to traverse this inner CSM and emerge into the
hypothesised larger-radii, lower-density CSM.
If this scenario is correct, our radio observations at 36, 76, and 148\,d after explosion 
should therefore probe ongoing CSM interaction between the fastest-moving ejecta and
material lost from the star around 1, 2, and 4\,yr before explosion, respectively.

Following \citet{2014ApJ...797....2K} and \citet{2016ApJ...818..111K}, we 
adopt the models of \cite{1998ApJ...499..810C} to describe the radio flux from SN~2015G
at the observed epochs,
assuming that the radial density profile of the CSM goes as $\rho \propto r^{-2}$ at large radii.
We parameterise the energy distribution of the shocked electrons as a power law  in
the electron Lorentz factor with index $p$, $n_{e}(\gamma_{e}) \propto \gamma_{e}^{-p}$;
we assume that the fractional energy densities in the relativistic electrons and in the magnetic field
are equivalent (i.e., the shocked material is in equipartition), $\epsilon_{e} = \epsilon_{B}$;
and we assume that $\epsilon_e = 0.1$.  These models include both the effects of synchrotron self-absorption
and free-free absorption from the CSM.

\begin{figure*}
    \includegraphics[width=0.45\textwidth]{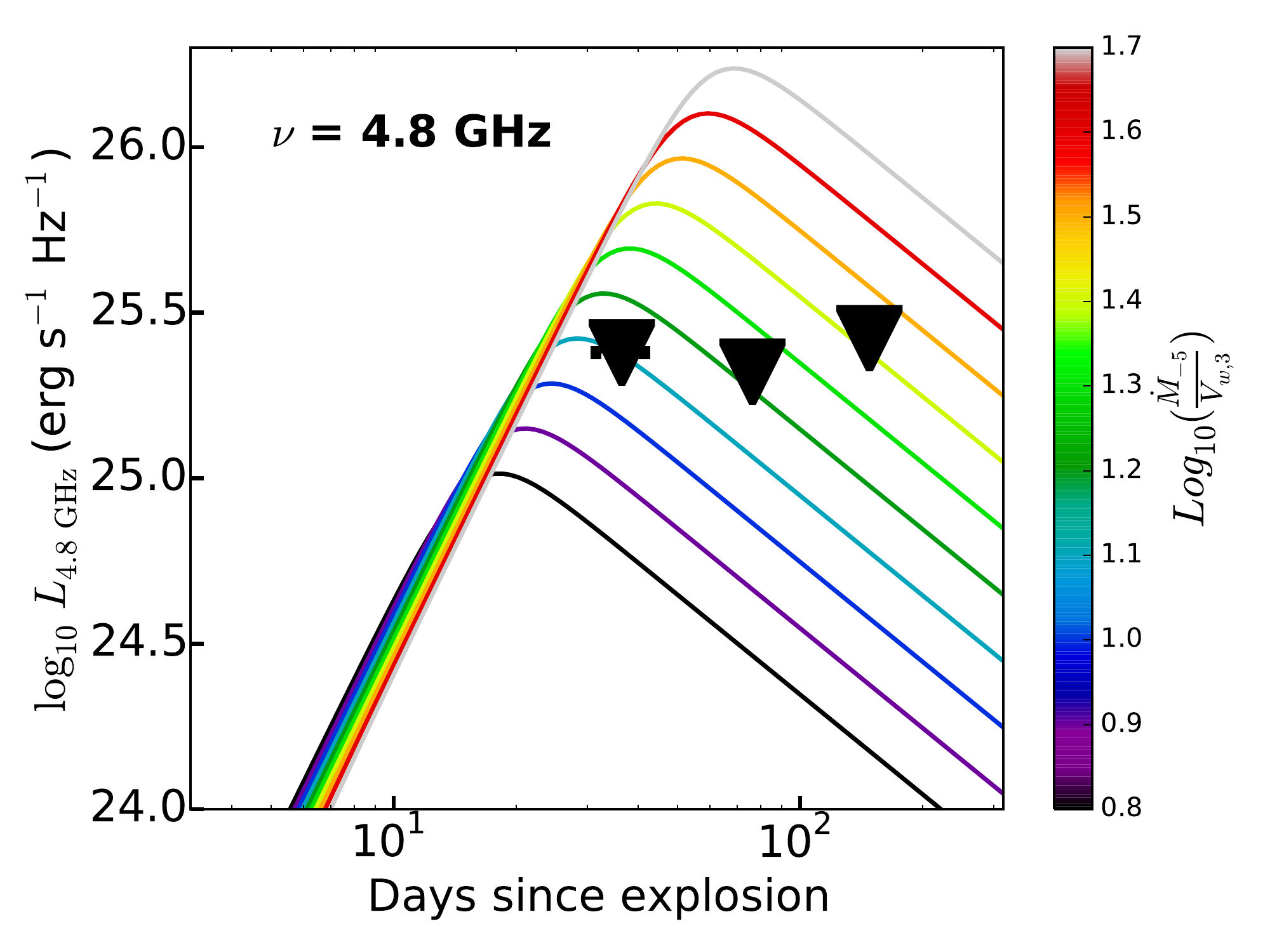}
    \includegraphics[width=0.45\textwidth]{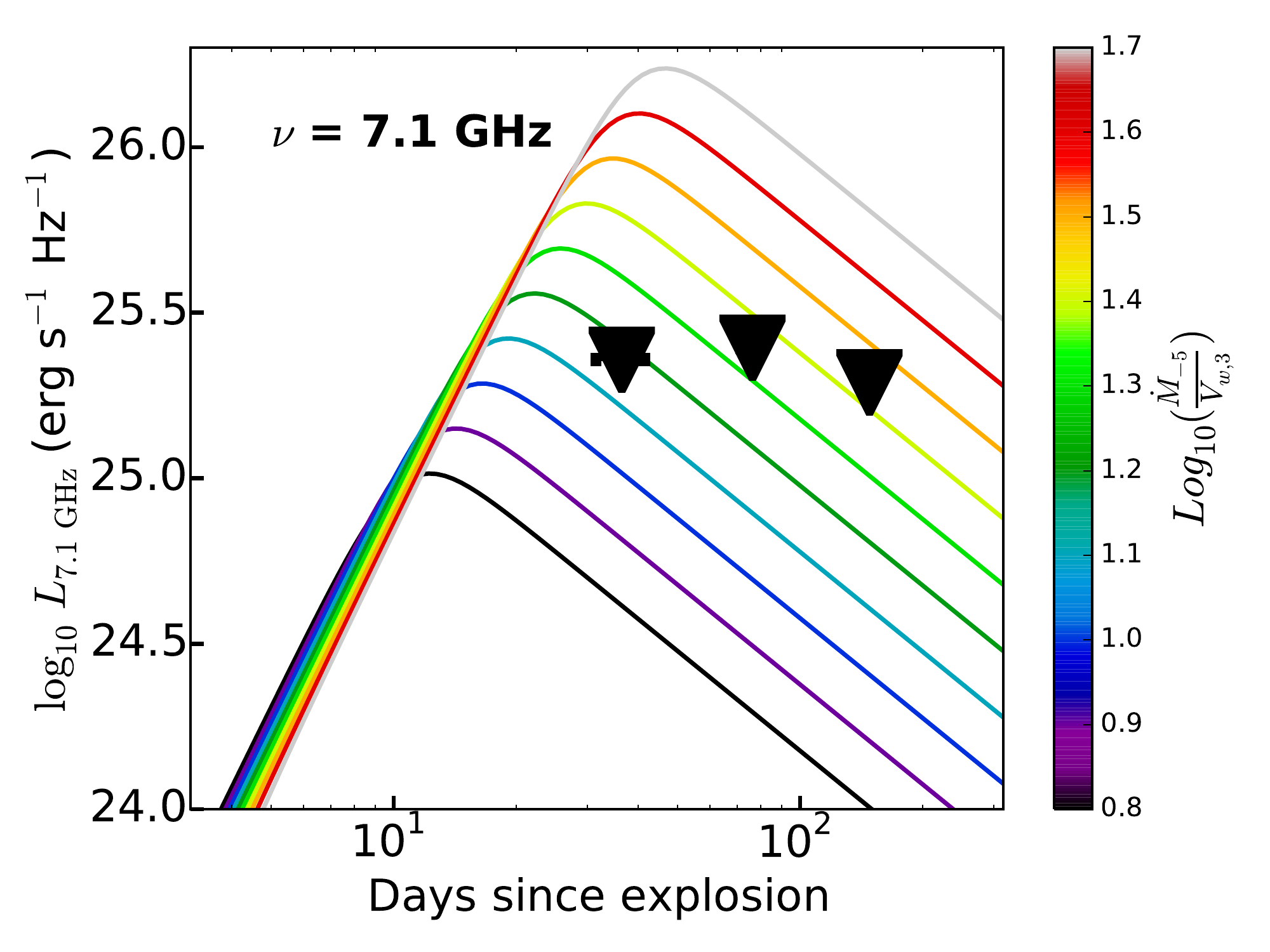}
    \caption{Our $3\sigma$ nondetections of SN~2015G at 4.8 and 7.1\,GHz, plotted
        against modeled light curves assuming a range of values for the wind parameter.  
        These models were constructed assuming $v_{\rm shock} = 10,000$\,\kms and $v_{\rm wind} = 1000$\,\kms.
        We adopt an explosion date of MJD 57080, 
       and we show horizontal error bars of $\pm 5$\,d to illustrate our uncertainty about the exact date of explosion
        (the plotted data points are often wider than those error bars).
    } \label{fig:radio}
\end{figure*}

In Figure~\ref{fig:radio} we plot our $3\sigma$ nondetections against modeled radio light curves
assuming $v_{\rm wind} = 1000$\,\kms\ and $v_{\rm shock} = 10,000$\,\kms.  
We show models for a range of values for the wind parameter, $10^{0.8} \leq A_{*} \leq 10^{1.7}$, where
$$A_{*} \equiv \frac{\dot{M} / 10^{-5} \, {\rm M}_{\odot} \, {\rm yr}^{-1}} {v_{\rm wind} / 10^{3} \, {\rm km\,s}^{-1}}$$
Our first epoch of observations produces the strongest constraint on the wind
mass-loss rate from SN~2015G's progenitor: $\dot{M} \lesssim 1 \times 10^{-4}\,M_\odot \,yr^{-1}$.
Assuming a slower wind velocity produces a more stringent constraint
(we find $\dot{M} \lesssim 10^{-5}\,{\rm M}_\odot \,{\rm yr}^{-1}$ if $v_{\rm wind} = 100$\,\kms), while assuming
a slower shock velocity relaxes the constraint
(we find $\dot{M} \lesssim 10^{-3}\,{\rm M}_\odot \,{\rm yr}^{-1}$ if $v_{\rm shock} = 5000$\,\kms).

Radio emission from SNe Ibn is still uncharted territory, and it is difficult to know
whether the assumptions (and therefore the models) outlined above are fully appropriate.  Similar
assumptions have been shown to be reasonable for stripped-envelope SNe with
detected radio light curves, but the diversity of radio signatures found for these events
is remarkable, especially among the SNe with evidence for unsteady pre-explosion mass loss from
their progenitor.  See, for example, PTF~11qcj, a radio-bright SN Ic that may have had
a SN~2006jc-like outburst from its progenitor $\sim 2$\,yr before core collapse \citep{2014ApJ...782...42C},
or SN~2014C, a SN Ib that began to interact with an H-rich dense shell a year after 
explosion and showed extreme variability in its radio light curve \citep{2017ApJ...835..140M}.

Our radio flux limits and the resultant CSM density limits shown in Figure~\ref{fig:radio} are surprising,
given the strong signatures of a dense CSM at low radii --- SNe~Ib/c with radio detections 
generally have peak luminosities in the range $10^{26}$--$10^{28}$\,erg\,s$^{-1}$\,Hz$^{-1}$ at these frequencies
\citep[e.g.,][]{2007AIPC..937..492S}.
If the CSM around SN~2015G had a structured density profile at large radii, 
perhaps with shells of material created by episodes of eruptive mass loss rather than
the steady wind-driven profile we model above, 
the radio emission powered by ongoing interaction could be entirely obscured
via free-free absorption within the CSM exterior to the shock.
Type IIn SNe in very dense environments, for example, sometimes exhibit
radio light curves with rise times of $\sim 1000$\,d because of this effect \citep[e.g.,][]{2012ApJ...755..110C,2015ApJ...810...32C}.
These SNe sustain the optical signatures of ongoing interaction, however, while SN~2015G's
narrow spectral features disappear and its interaction-powered optical brightness fades away ---
differences which argue that the radial profile of the CSM surrounding SN~2015G must be 
quite dissimilar from the (relatively) smooth density profiles inferred for the SNe~IIn above.
A further worry is that our models do not account for any CSM clumpiness or global asymmetry, though our
optical observations argue that the SN~2015G system is strongly asymmetric.
Without clear detections and lacking a true radio light curve, many uncertainties remain.

\subsection{Progenitor Constraints}
\label{sec:progenitor}

\begin{figure*}
  \begin{minipage}[c]{0.7\textwidth}
    \includegraphics[width=\textwidth]{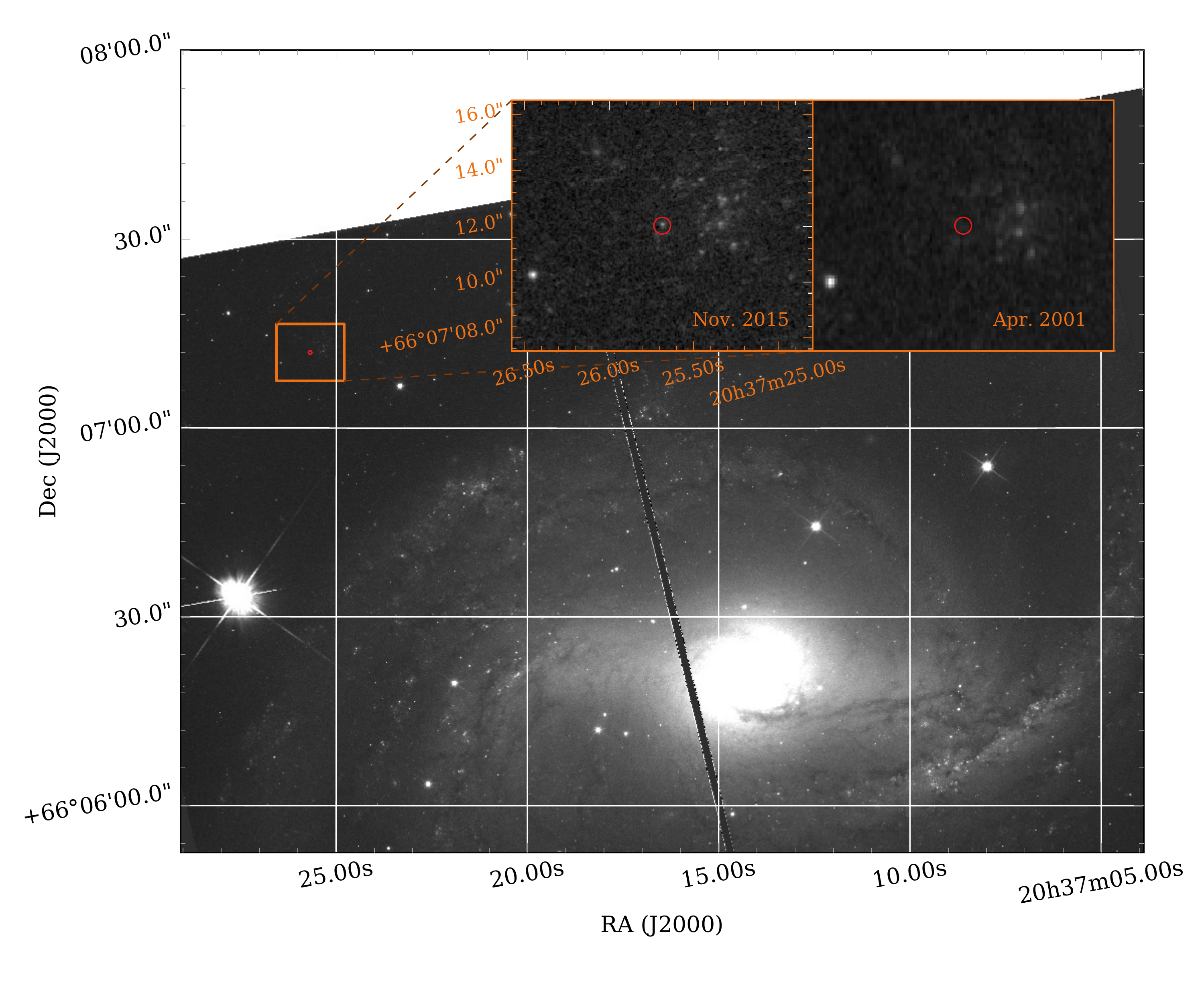}
  \end{minipage}\hfill
  \begin{minipage}[c]{0.27\textwidth}
    \caption{{\it HST} image of SN~2015G taken $\sim 8$\,months after discovery
    and a pre-explosion image of the SN site from 2001, both observed
    through the {\it F814W} filter.
    The main frame shows the large offset between SN~2015G and its host, 
    while the two inset frames display close-up views of the fading SN (left)
    and the pre-explosion progenitor nondetection (right).
    We indicate the location of the SN using a red circle with
    a radius of $0.3''$ (larger than our $3\sigma$ uncertainty in 
   the SN location).
    } \label{fig:finder}
  \end{minipage}
\end{figure*}

SN~2015G is one of the nearest known SNe~Ibn, and as such it provides us with a unique opportunity to study
the progenitor star and local environment of a member of this rare subclass. 
A preliminary search for the progenitor in the {\it HST}/WFPC2 images from 2001 was presented by \citet{2015ATel.7563....1M}.
They established the SN position in these archival pre-SN data using a ground-based $I$-band image and did not detect a progenitor
candidate there. Neglecting any extinction within the host galaxy, they estimate upper limits on the
luminosity of a progenitor at $M_I > -6.4$ mag and $M_V > -7.1$ mag. 

We initially used our {\it HST} Target of Opportunity (ToO) WFC3 images from GO-13683 to provide
a better position for the SN in the 2001 WFPC2 {\it F555W} data. However, since the individual
frame times for the ToO observations were only 10\,s each (for a total of 240\,s) and we observed
in subarray mode, there were only 7 stars in common between the two image datasets. Consequently,
we could only register the images with a $1\sigma$ uncertainty of 0.61 WF pixel
(the SN site is on the WF2 chip of the WFPC2 array). We therefore registered the pre-SN images
to the much deeper WFC3 full-array data from GO-14149 (total exposure times of 780\,s in {\it F555W},
710\,s in {\it F814W}). We found 30 star-like objects in common between the images and were able to
achieve an astrometric registration that was somewhat better, with a $1\sigma$ uncertainty of
0.38 WF pixel. We note that the positions for the SN in the pre-SN data estimated from the two
different WFC3 image sets agree to 0.49 pixel. We also do not see a progenitor candidate at this
position, nor does {\tt DOLPHOT} detect any object there. In Figure~\ref{fig:finder} we present the
{\it HST}/WFC3 image of SN 2015G from November 2015, a close-up view of the SN and its local environment,
and a close-up view of the progenitor nondetection from 2001 (all in the {\it F814W} band). 

As Figure~\ref{fig:finder} shows, SN~2015G exploded far from the
bulk of the stellar mass in NGC~6951 and far from the major star-forming regions.
As noted by \citet{2015ATel.7563....1M},
however, there is a small but conspicuous clump of bright and blue stars near that location.
The centre of this clump is $\sim 2''$ west of SN~2015G, a distance of $\sim 200$\,pc.
Most of the stars in the clump are within $\sim 100$\,pc of the centre --- if SN~2015G's progenitor
formed as a part of this clump, it appears to have traveled 
an appreciable distance from its birthplace. Alternatively, the progenitor may have
formed within a smaller stellar subgroup, possibly at a different time.

Our final spectrum of SN~2015G was observed with this cluster
along the slit, and a narrow \ion{H}{1} line arising from this small star-forming region 
was detected.  The redshift as measured from this emission line is in good agreement
with the published redshift of the host galaxy, with an observed wavelength
of $6592.33 \pm 0.06$\,\AA\ (as measured via maximum-likelihood MCMC fitting,
assuming a Gaussian line profile). This implies a 
redshift of $0.00450$, or a line-of-sight velocity within 100\,\kms\      
of that of NGC~6951.

We attempted artificial-star tests with {\tt DOLPHOT} on the images
from 2001, injecting an artificial star at the exact SN position, and found the following
nondetection upper limits: {\it F555W} $\gtrsim 26.7$\,mag and {\it F814w} $\gtrsim$ 25.4\,mag.
These are consistent with the formal $3\sigma$ source detections by {\tt DOLPHOT} at the SN's location,
and they translate into absolute upper limits of
${\rm M}_{F555W} \gtrsim -6.4$\,mag and
${\rm M}_{F814W} \gtrsim -7.1$ mag.
These limits are essentially the same as those found by \citet{2015ATel.7563....1M}, though
their assumptions of the distance and reddening to the SN differed from ours.

Assuming the progenitor of SN~2015G was a single supergiant, and that it exploded at
the terminus of its evolutionary track as something other than a hydrogen-rich supergiant,
we compared our detection upper limits with the MESA Isochrones and Stellar Tracks
\citep[MIST;][]{2011ApJS..192....3P,2013ApJS..208....4P,2015ApJS..220...15P,2016ApJ...823..102C}
at solar and slightly subsolar ([Fe/H] $= -0.25$) metallicities, adjusted for the distance
and reddening to SN~2015G as described above.
We find the following limits on the zero-age main sequence mass
by comparison with our {\it F555W} limit (the more constraining of the two):
$M_0 \lesssim 18$\,M$_{\odot}$ (solar) and $M_0 \lesssim 20$\,M$_{\odot}$ (subsolar).
These limits effectively rule out a single massive ($M_0 \gtrsim 30$\,M$_{\odot}$) Wolf-Rayet star progenitor
(assuming that the luminosity at the time of collapse is similar to the luminosities observed for Wolf-Rayet stars in the MW).
Comparisons, instead, with the interacting binary models of \citet{2015ApJ...809..131K} show that some
configurations are disallowed by our upper limits but a variety of possible binary models remain viable.
Note that {\it HST} images of the field were obtained in 1994 through the {\it F218W} and {\it F547M} filters,
but they do not provide deeper constraints.

\citet{2016ApJ...833..128M} have recently reported the likely detection of a binary
companion to the progenitor of SN~2006jc in {\it HST} images obtained 4\,yr after the SN explosion.
The pre-explosion upper limits we find for the progenitor system of SN~2015G are marginally
consistent with a similarly luminous source, depending upon the assumed
properties of the extinction along the line of sight within the host galaxy.

\begin{figure}
  \includegraphics[width=0.9\columnwidth]{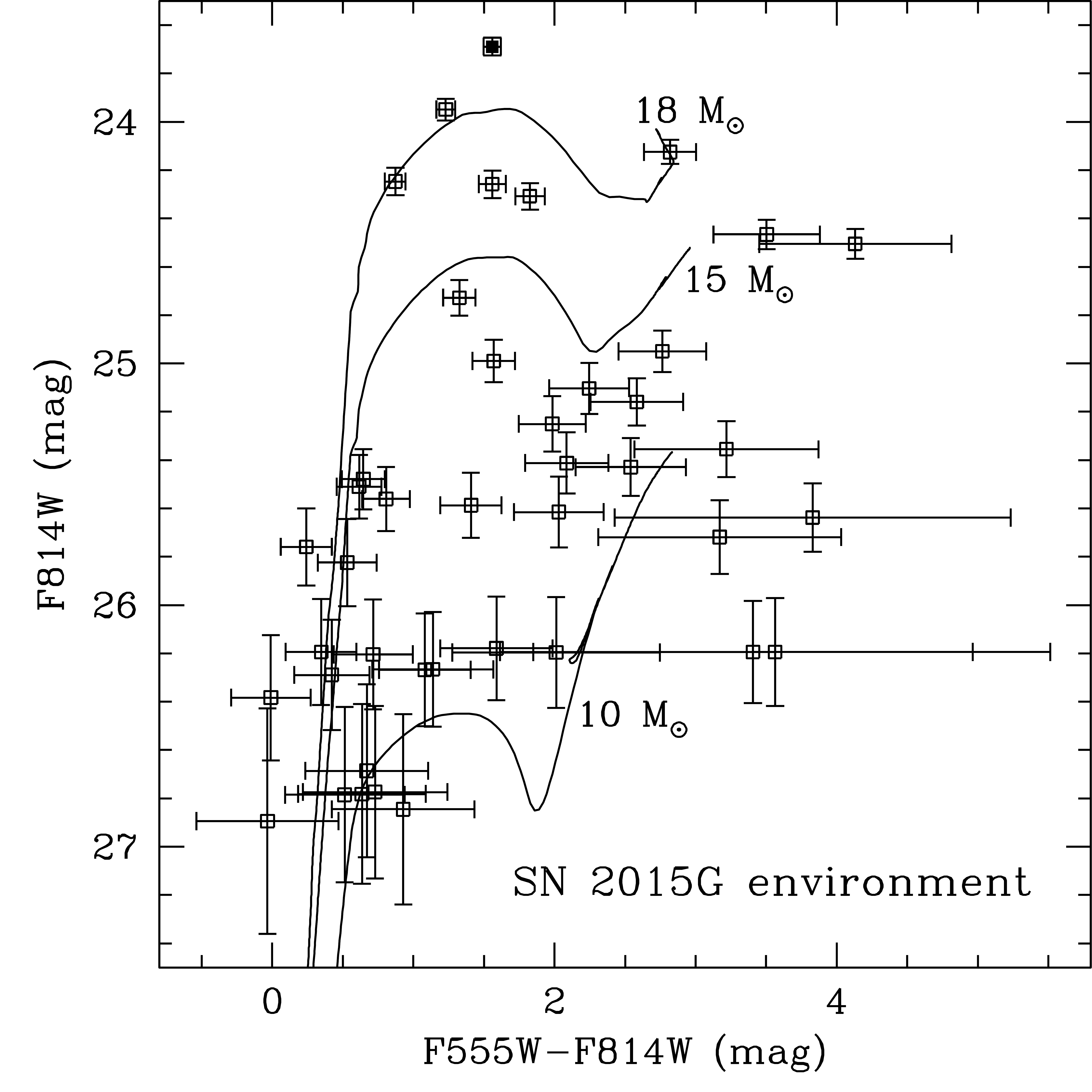}
  \caption{Colour-magnitude diagram of star-like objects (open squares) detected by {\tt DOLPHOT}
           within a 43-pixel ($\sim 200$\,pc) radius of SN~2015G's position in {\it HST}/WFC3 images
           obtained in 2015 by program GO-14149. The SN itself is shown as a filled square.
           For comparison we plot MIST stellar evolutionary tracks at slightly subsolar ([Fe/H] $= -0.25$)
           metallicity and initial masses of 10, 15, and 18\,M$_{\odot}$, adjusted to the assumed
           distance and reddening of SN~2015G.
  \label{fig:environ} }
\end{figure}

We also considered the stellar environment of SN~2015G to constrain the properties of the progenitor.
We analyzed objects within a 43-pixel ($\sim 200$\,pc) radius of the SN detected by {\tt DOLPHOT} in the WFC3
{\it F555W} and {\it F814W} images from GO-14149 that had a {\tt DOLPHOT} object type of ``1'' (i.e., star-like).
The resulting colour-magnitude diagram is shown in Figure~\ref{fig:environ}. We again compared the stellar photometry to the
MIST tracks and found that, assuming that the population of stars in this region are coeval and that the
SN progenitor itself was not rejuvenated in its evolution as a result of binary interaction, the highest
initial mass the progenitor star could have had was $\sim 18$\,M$_{\odot}$, consistent with the upper limits
calculated from our progenitor nondetection.

Comparing our Figure~\ref{fig:finder} to Figure~1 of \citet{2016ApJ...833..128M},
we note the remarkable similarity between SN~2006jc's local environment and 
that of SN~2015G: both SNe exploded in sparse areas of their hosts near 
clumps of young, massive stars but offset from them by $\gtrsim 100$\,pc.
The Type IIn SN 2009ip was also quite isolated \citep{2016MNRAS.463.2904S},
and \citet{2015MNRAS.447..598S} show that luminous blue variables
(LBVs) in the MW, often proposed to be Galactic analogues for the
progenitors of strongly interacting SNe, are as well.
They interpret the isolation of LBVs as evidence that
they are mass gainers in binary pairs which get rejuvenated by mass
exchange and receive a kick when their (more massive) companion
explodes, allowing them to travel far from their birth sites before their own deaths
\citep[note that these results are under some debate; e.g.,][]{2016ApJ...825...64H}.

The presence of a dense CSM surrounding SN~2015G
suggests a recent history of extreme mass loss from,
and therefore variability of, the progenitor star.
An LBV-like bright outburst from SN~Ibn 2006jc's progenitor
was observed some 2\,yr before the SN itself 
\citep[$M_r \approx -14.1$\,mag;][]{2006CBET..666....1N,2007ApJ...657L.105F,2007Natur.447..829P}.
There have also been a few SN~IIn progenitors detected in outburst in the years prior to core collapse
\citep[e.g.,][]{2013ApJ...779L...8F,2013MNRAS.430.1801M,2013ApJ...767....1P,2014ApJ...780...21M,2014ARA&A..52..487S,2014ApJ...789..104O,2016MNRAS.463.3894E,2017A&A...599A.129T}, 
though similar outbursts have, in other cases, been ruled out
 \citep[e.g.,][]{2015MNRAS.450..246B}. KAIT has been monitoring NGC~6951 for almost
20\,yr and we searched this extensive dataset for evidence of pre-explosion variability.

\begin{figure}
  \includegraphics[width=\columnwidth]{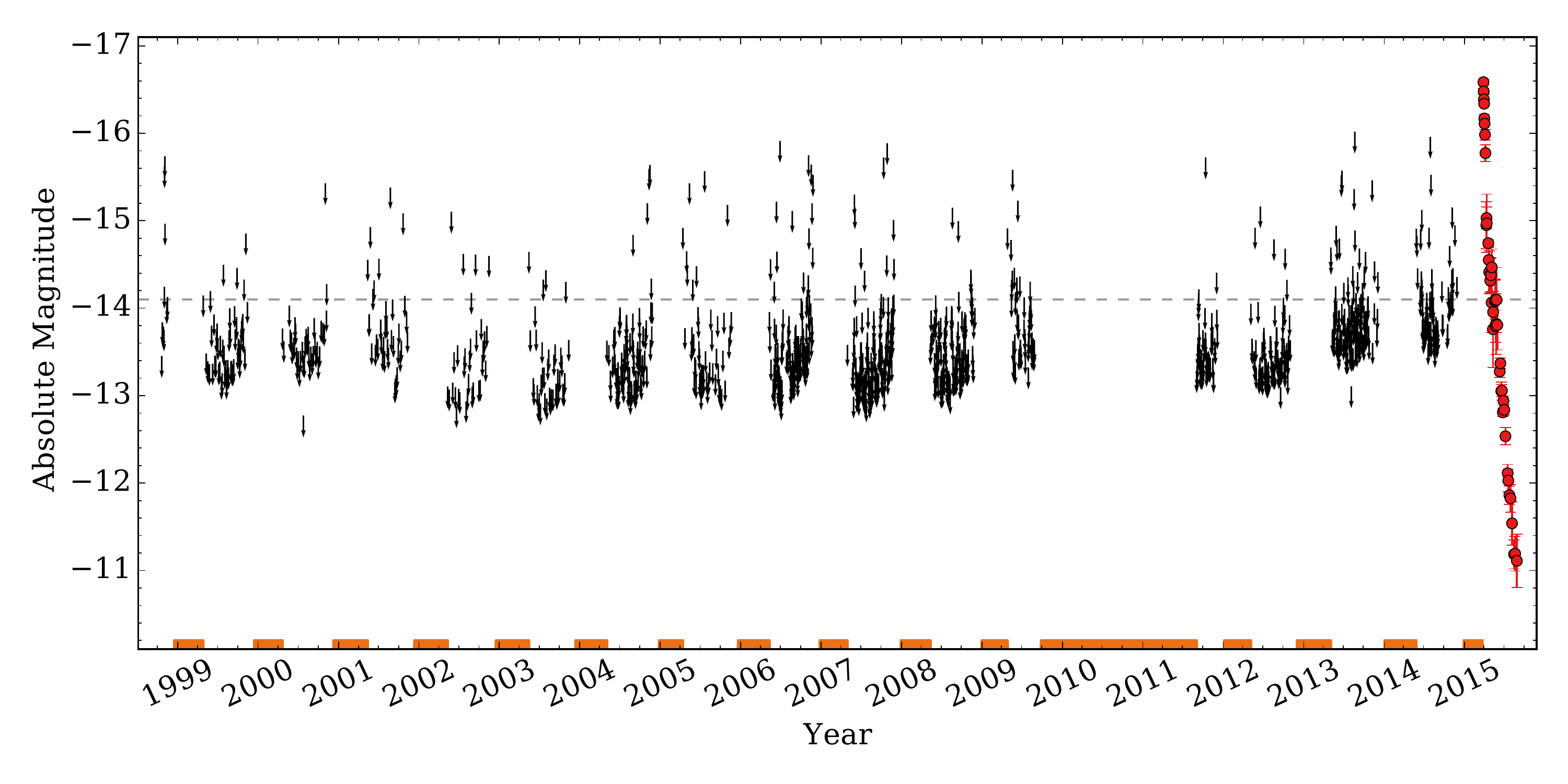}
  \caption{Our 1$\sigma$ pre-explosion nondetections from KAIT unfiltered images.
    The $R$-band light curve of SN~2015G is shown at the far right.
    Timespans for which no upper limit had been obtained for at least 1\,month
    are marked in orange along the bottom, and a dashed line indicates the 
    absolute magnitude of SN~2006jc's pre-explosion outburst \citep{2007ApJ...657L.105F}.
  \label{fig:uplims} }
\end{figure}

Examining 1248 unfiltered images taken between 1998 and 2015
with detectability thresholds deeper than 17.0\,mag,
we find no detections at the SN location and no evidence for previous outbursts of
the progenitor brighter than $-13.3 \pm 0.5$\,mag
(median and standard deviation of the detection thresholds among all images).
Figure~\ref{fig:uplims} plots our $1\sigma$ nondetections and the observed light curve of SN~2015G,
with the luminosity of SN~2006jc's pre-explosion outburst indicated for comparison.
(The {\it HST} nondetections described above provide additional extremely strong
constraints in 2001 May, off the bottom of the scale of Figure~\ref{fig:uplims}.)
The SN field was inaccessible to our telescopes for several months every year, so there are significant gaps;
the orange bars along the bottom of Figure~\ref{fig:uplims} mark every night on which more
than 1\,month had passed since the previous upper limit.
Approximately 45\% of the nights between October 1998 and the SN discovery in March 2015
fall into such a gap.

Though these observations rule out any long-lasting 
luminous outbursts in the last 20\,yr, the outburst from
SN~2006jc's progenitor was observed to fade rapidly after
discovery \citep[$\sim 0.16$\,mag\,d$^{-1}$ over the 9\,d it was detected,][]{2007Natur.447..829P}; thus, our nondetections argue neither for nor strongly
against a SN~2006jc-like outburst from SN~2015G's progenitor
\citep{2006CBET..666....1N,2007Natur.447..829P}.

\subsection{A Rough Schematic of the SN~2015G System}
\label{sec:system}

Based upon the above observations, 
we interpret SN~2015G to be a Type Ib SN explosion
modified by additional
luminosity arising via the collision between explosive ejecta and dense CSM,
with the collision between SN ejecta and the CSM converting the
kinetic energy of the ejecta into radiative luminosity
\citep[e.g.,][]{1994MNRAS.268..173C,1994ApJ...420..268C,2011ApJ...729L...6C}.

The luminous yet rapidly fading light-curve peak 
settles into a slower decline rate, while either
ongoing (weaker) interaction or a (relatively small amount of) radioactive
material powers the luminosity of the late-time light-curve tail.
Our spectroscopic monitoring of SN~2015G began after 
shock breakout and peak luminosity, and
the early-time spectra of the event show a cooling blue continuum topped
by relatively broad emission lines (arising from the ejecta and
perhaps the shocked and accelerated CSM)
and narrow P-Cygni lines (arising from the unshocked
and extended CSM at larger radii).
The spectral lines at early phases are centred at a 
velocity of 0\,\kms, and therefore the CSM in which these
early lines formed likely exhibited a range of velocity vectors
more or less symmetrically distributed around the progenitor.

These narrow P-Cygni features disappeared from our spectra
$\sim 10$\,d after the discovery of SN~2015G,
or $\sim 35$\,d after our roughly estimated explosion date.
Coupled with our radio nondetections (the first of which was observed 36\,d post-explosion),
this argues that the dense CSM was predominantly located at small radii and
was therefore likely lost from the surface of the progenitor in the last year or so
before core collapse.

As the light curve settles into its late-time decline rate,
the broader features transition
from pure emission into a P-Cygni profile,
likely arising from some mixture of the swept-up CSM and the ejecta.
The light curve then continues to decline steadily as the
continuum, and therefore the absorption features, fade away. 
As the ejecta and accelerated CSM continue to expand and
the density drops, forbidden emission lines become prominent.
The evolution of all emission lines redward argues that the 
line flux at late times arises predominantly within receding material, unlike
the early emission-line flux.

Whether ongoing weak CSM interaction or a relatively small amount
of $^{56}$Ni powers the late-time light curves of SNe Ibn is still a difficult question.
SN~2015G's late-time decline at redder wavelengths (the {\it I} and {\it F814W} passbands)
appears to be very similar to that at bluer wavelengths ({\it V} and {\it F555W}), arguing
that the blue pseudocontinuum and the (mostly red) emission lines are powered
by the same process. The blue pseudocontinuum is generally understood to be
powered via CSM interaction, and so this argues that the line emission also arises
from CSM interaction.

In contrast, the linear decline of the late-time light curve and the homogeneity of
light-curve shapes among SNe~Ibn argue for the radioactively powered interpretation.
If SN~Ibn light curves are interaction-powered on the tail, the diversity of late-time light-curve 
properties should reflect the diversity of CSM configurations around the progenitors;
it would be surprising if these CSM configurations (and therefore the progenitor pre-explosion mass-loss histories)
were so similar across different SNe \citep[e.g.,][]{2008MNRAS.389..113P,2017ApJ...836..158H}.
The light curves of SNe IIn (which assuredly are powered largely by interaction) are very heterogeneous,
as expected \citep[e.g.,][]{2012ApJ...744...10K},
though comparisons with hydrogen-rich SNe must be made with caution;
the lack of hydrogen in SNe~Ibn may force the continuum opacity significantly lower.
Note that the late-time luminosities of SNe Ibn are low compared to those of normal SNe~Ib/Ic ---
if they are radioactively powered at late times, it seems they must produce a relatively small amount of $^{56}$Ni.

The systemic redshift of the \ion{Ca}{2} and \ion{He}{1} lines
implies a severe (and peculiar) asymmetry of the system, likely due
to an asymmetry of the CSM with which SN~2015G's ejecta are 
interacting at these phases (assuming these lines are interaction-powered). 
However, we do observe some polarization intrinsic to SN~2015G at early times,
and a less-than-spherical explosion itself may also be playing a role.
Not only are asymmetric geometries often invoked to understand
the observed properties of core-collapse SNe
\citep[e.g.,][]{2005Sci...308.1284M,2008Sci...319.1220M,2008ApJ...687L...9M,2009MNRAS.397..677T,2010ApJ...709.1343M},
but both the analysis of some SN remnants and the results of 
modern 2- and 3-dimensional modeling efforts of the core-collapse mechanism
itself argue that asymmetric (sometimes unipolar) explosions are possible and may even be common.

The Puppis A SN remnant \citep{1996ApJ...465L..43P}, for example,
shows a compact neutron-star remnant with a systemic velocity of some 1000\,\kms\ 
along a vector opposite that of the bulk ejecta velocity, arguing that the neutron star
received a substantive kick from the core-collapse event and that the ejecta received
a similar kick in the opposite direction.
Explosion asymmetries of lesser degree have also been observed 
\citep[e.g., the so-called ``Bochum'' event of SN~1987A;][]{1989PASP..101..137P}.
From the modeling side, several teams have shown that low-order spherical harmonics
of the exploding core may well manifest themselves in large-scale
asymmetries of the explosion
\citep[e.g.,][]{2010PASJ...62L..49S,2012ApJ...755..138H,2014AAS...22321605C,2014ApJ...785..123C}.

\section{Conclusion}
\label{sec:conclusion}

SN~2015G, which exploded in NGC~6951 at a distance of 23.2\,Mpc,
is one of the nearest known SNe~Ibn. Though it was discovered after peak
brightness, we have been able to accrue a remarkable dataset on this event, 
making it one of the best-studied SNe of this rare type and highlighting
both strong similarities with and differences from the archetypical SN~Ibn~2006jc.

\citet{2017ApJ...836..158H} argue for two spectroscopically defined
subclasses of SNe Ibn, but our observations of SN~2015G show
that it exhibited properties of both proposed subclasses.
Rather than two physically distinct subclasses, perhaps a continuum of CSM properties
surrounding the SN produces a continuum of spectroscopic properties; 
this question should be investigated further as more SNe~Ibn are identified and studied.

Archival {\it HST} images of the resolved SN explosion site argue against a single massive WR-like
progenitor for SN~2015G.  Given the recent likely detection of a binary companion
to SN~2006jc's progenitor, the isolation of SN~2015G's explosion site and the
well-determined position of the SN in multiple {\it HST} images makes SN~2015G an
excellent candidate for a similar study in the future.

The data presented here argue that extreme mass loss from the progenitor of SN~2015G occurred
soon ($\sim 1$\,yr) before core collapse, and that the SN~2015G system was asymmetric.
Asymmetries in stripped-envelope SNe are common, but the degree of asymmetry shown by
the late-time spectra of SN~2015G has not been observed in a SN~Ibn before now.
A dedicated effort to obtain more high-resolution spectra and better late-time coverage
of SNe~Ibn is called for to understand whether severe asymmetry is characteristic of SNe~Ibn
or a unique trait of the SN~2015G system.

\section*{Acknowledgements}

We thank K.~Shen, C.~Harris, and J.~Schwab for helpful discussions.
We thank our referee, A.~Pastorello, for his productive comments, which
have improved this paper.
A.V.F.'s supernova group at U.C. Berkeley is supported by US National
Science Foundation (NSF) grant AST-1211916, Gary \& Cynthia Bengier,
the Richard \& Rhoda Goldman Fund, the Christopher R. Redlich Fund,
the TABASGO Foundation, and the Miller Institute for Basic Research
in Science (U.C. Berkeley).
His work was conducted in part at the Aspen Center for Physics,
which is supported by NSF grant PHY-1607611; he thanks the Center for
its hospitality during the neutron stars workshop in June and July 2017.
Support for {\it HST} programs GO-13683,
GO-13797, GO-14149, AR-14295, and GO-14668 was provided by the National
Aeronautics and Space Administration (NASA) through grants from the
Space Telescope Science Institute, which is operated by the
Association of Universities for Research in Astronomy (AURA), Inc.,
under NASA contract NAS5-26555.
The UCSC group is supported in part by NSF grant AST-1518052 and from
fellowships from the Alfred P.\ Sloan Foundation and the David and
Lucile Packard Foundation to R.J.F.

Some of the data presented herein were obtained at the W. M. Keck
Observatory, which is operated as a scientific partnership among the
California Institute of Technology, the University of California, and
NASA; the observatory was made possible by the generous financial
support of the W. M. Keck Foundation.  KAIT and its ongoing operation
were made possible by donations from Sun Microsystems, Inc., the
Hewlett-Packard Company, AutoScope Corporation, Lick Observatory, the
NSF, the University of California, the Sylvia and Jim Katzman
Foundation, and the TABASGO Foundation. Research at Lick Observatory
is partially supported by a generous gift from Google.  We are
grateful to the staffs of Lick, Keck, and the other observatories
where we obtained data for their excellent assistance.

This work made use of {\it Swift}/UVOT data reduced by P. J. Brown for
the {\it Swift} Optical/Ultraviolet Supernova Archive (SOUSA).  SOUSA
is supported by NASA's Astrophysics Data Analysis Program through
grant NNX13AF35G.  This research has made use of data and/or software
provided by the High Energy Astrophysics Science Archive Research
Center (HEASARC), which is a service of the Astrophysics Science
Division at NASA/GSFC and the High Energy Astrophysics Division of the
Smithsonian Astrophysical Observatory.  This research has made use of
the NASA/IPAC Extragalactic Database (NED) which is operated by the
Jet Propulsion Laboratory, California Institute of Technology, under
contract with NASA.
s



\bibliographystyle{mnras}
\bibliography{bib}



\bsp	
\label{lastpage}
\end{document}

%% file: phottable.tex
\begin{table*}
\centering
\small
\caption{Table of Photometric Observations}
\label{tab:photometry}
\begin{tabular}{ c c c | c c c }
\hline
Date & MJD & Phase                 & Magnitude & Passband & Telescope \\
(UT) &     & (days) &           &          &           \\
\hline
2015-03-27.51 & 57108.51 & 3.74 & 17.07$\pm$0.07 & B & KAIT \\ 
2015-03-27.51 & 57108.51 & 3.74 & 16.61$\pm$0.04 & V & KAIT \\ 
2015-03-27.51 & 57108.51 & 3.74 & 16.20$\pm$0.03 & R & KAIT \\ 
2015-03-27.51 & 57108.51 & 3.74 & 15.81$\pm$0.04 & I & KAIT \\ 
2015-03-28.53 & 57109.53 & 4.75 & 17.17$\pm$0.08 & B & KAIT \\ 
2015-03-28.53 & 57109.53 & 4.75 & 16.71$\pm$0.04 & V & KAIT \\ 
2015-03-28.53 & 57109.53 & 4.75 & 16.30$\pm$0.03 & R & KAIT \\ 
2015-03-28.53 & 57109.53 & 4.75 & 15.86$\pm$0.04 & I & KAIT \\ 
2015-03-29.54 & 57110.54 & 5.76 & 17.25$\pm$0.15 & B & KAIT \\ 
2015-03-29.54 & 57110.54 & 5.76 & 16.77$\pm$0.08 & V & KAIT \\ 
2015-03-29.54 & 57110.54 & 5.77 & 16.40$\pm$0.05 & R & KAIT \\ 
2015-03-29.54 & 57110.54 & 5.77 & 15.94$\pm$0.08 & I & KAIT \\ 
2015-03-30.52 & 57111.52 & 6.74 & 17.63$\pm$0.02 & B & Nickel \\ 
2015-03-30.52 & 57111.52 & 6.74 & 16.94$\pm$0.01 & V & Nickel \\ 
2015-03-30.53 & 57111.53 & 6.75 & 16.44$\pm$0.01 & R & Nickel \\ 
2015-03-30.53 & 57111.53 & 6.75 & 15.99$\pm$0.01 & I & Nickel \\ 
2015-03-31.53 & 57112.53 & 7.76 & 17.63$\pm$0.21 & B & KAIT \\ 
2015-03-31.54 & 57112.54 & 7.76 & 17.10$\pm$0.07 & V & KAIT \\ 
2015-03-31.54 & 57112.54 & 7.76 & 16.61$\pm$0.05 & R & KAIT \\ 
2015-03-31.54 & 57112.54 & 7.76 & 16.09$\pm$0.06 & I & KAIT \\ 
\multicolumn{4}{c}{ {\it Truncated; full table available digitally.}} \\
\hline
\end{tabular}
\end{table*}

%% file: speclog.tex
\begin{table*}
\centering
\small
\caption{Journal of Spectroscopic Observations}
\label{tab:speclog}
\begin{tabular}{ c c c | c c c }
\hline
Date & MJD & Phase  & Tel./Inst. & Wavelength & Resolution \\
(UT) &     & (days) &            & (\AA)      & (\AA)      \\
\hline
2015-03-26.51 & 57107.51 & 2.73 & Shane/Kast & 3,450--10,860 & 10 \\ 
2015-03-27.50 & 57108.50 & 3.72 & Bok/B\&C & 6,190--7,350 & 2 \\ 
2015-03-27.44 & 57108.44 & 3.66 & Bok/B\&C & 4,400--8,170 & 10 \\ 
2015-03-27.50 & 57108.50 & 3.72 & Shane/Kast & 3,450--10,880 & 10 \\ 
2015-03-28.46 & 57109.46 & 4.68 & Shane/Kast & 3,460--10,850 & 10 \\ 
2015-03-28.48 & 57109.48 & 4.70 & Shane/Kast & 3,450--7,960 & 6 \\ 
2015-03-28.50 & 57109.50 & 4.72 & Shane/Kast (SpecPol) & 4,550--9,940 & 10 \\ 
2015-04-04.64 & 57116.64 & 11.86 & HST/STIS/MAMA & 1,570--10,230 & 10 \\ 
2015-04-10.49 & 57122.49 & 17.71 & Bok/B\&C & 5,700--6,870 & 2 \\ 
2015-04-10.43 & 57122.43 & 17.65 & Bok/B\&C & 4,400--8,170 & 10 \\ 
2015-04-10.21 & 57122.21 & 17.43 & Liverpool/SPRAT & 4,010--7,950 & 18 \\ 
2015-04-11.50 & 57123.50 & 18.72 & HST/STIS/MAMA & 1,570--10,230 & 10 \\ 
2015-04-12.16 & 57124.16 & 19.38 & Liverpool/SPRAT & 4,010--7,950 & 18 \\ 
2015-04-15.44 & 57127.44 & 22.66 & LBT/MODS & 3,500--10,000 & 7 \\ 
2015-04-16.49 & 57128.49 & 23.71 & Shane/Kast & 3,450--10,860 & 10 \\ 
2015-04-18.19 & 57130.19 & 25.41 & Liverpool/SPRAT & 4,010--7,950 & 18 \\ 
2015-04-19.52 & 57131.52 & 26.74 & Shane/Kast & 5,000--10,000 & 10 \\ 
2015-04-20.31 & 57132.31 & 27.53 & HST/STIS/MAMA & 1,570--10,230 & 10 \\ 
2015-04-27.43 & 57139.43 & 34.65 & Shane/Kast & 3,500--10,500 & 10 \\ 
2015-04-28.43 & 57140.43 & 35.65 & Bok/B\&C & 4,400--8,060 & 10 \\ 
2015-04-28.09 & 57140.09 & 35.31 & Liverpool/SPRAT & 4,000--7,950 & 18 \\ 
2015-04-30.38 & 57142.38 & 37.60 & MMT/Blue Channel & 5,710--7,000 & 2 \\ 
2015-04-30.12 & 57142.12 & 37.34 & Liverpool/SPRAT & 4,020--7,990 & 18 \\ 
2015-05-02.15 & 57144.15 & 39.37 & Liverpool/SPRAT & 4,020--7,990 & 18 \\ 
2015-05-06.12 & 57148.12 & 43.34 & Liverpool/SPRAT & 4,020--7,990 & 18 \\ 
2015-05-09.09 & 57151.09 & 46.31 & Liverpool/SPRAT & 4,020--7,990 & 18 \\ 
2015-05-11.18 & 57153.18 & 48.40 & Liverpool/SPRAT & 4,020--7,990 & 18 \\ 
2015-05-14.17 & 57156.17 & 51.39 & Liverpool/SPRAT & 4,000--7,960 & 18 \\ 
2015-05-20.56 & 57162.56 & 57.78 & Keck-2/DEIMOS & 4,410--9,640 & 4 \\ 
2015-05-26.48 & 57168.48 & 63.70 & Shane/Kast & 3,460--10,880 & 10 \\ 
2015-06-22.43 & 57195.43 & 90.65 & Shane/Kast & 3,440--10,860 & 10 \\ 
2015-06-23.47 & 57196.47 & 91.69 & Shane/Kast & 3,450--10,850 & 10 \\ 
2015-06-24.46 & 57197.46 & 92.68 & Shane/Kast & 3,440--10,860 & 10 \\ 
2015-07-16.53 & 57219.53 & 114.75 & Keck-1/LRIS & 3,100--10,330 & 7 \\ 
2015-07-20.37 & 57223.37 & 118.59 & Shane/Kast & 3,430--10,820 & 10 \\ 
2015-08-12.32 & 57246.32 & 141.54 & Shane/Kast & 3,420--10,830 & 10 \\ 
2015-09-16.38 & 57281.38 & 176.60 & Keck-1/LRIS & 3,560--10,320 & 7 \\ 

\hline
\end{tabular}
\end{table*}